%% file: cvn.tex
\documentstyle[epsf,psfig]{l-aa}

\begin{document}

\newcommand{\as}[2]{$#1''\,\hspace{-1.7mm}.\hspace{.1mm}#2$}
\newcommand{\am}[2]{$#1'\,\hspace{-1.7mm}.\hspace{.0mm}#2$}
\newcommand{\dgr}{\mbox{$^\circ$}}    
\newcommand{\E}[1]{\mbox{${}\,10^{#1}{}$}}
\newcommand{\ea}{{\it et al.}}
\newcommand{\grd}[2]{\mbox{#1\fdg #2}}
\newcommand{\gsim}{\stackrel{>}{_{\sim}}}
\newcommand{\lsim}{\stackrel{<}{_{\sim}}}
\newcommand{\Ha}{\mbox{H$\alpha$}}
\newcommand{\HI}{\mbox{H\,{\sc i}}\,\,}
\newcommand{\HIshort}{\mbox{H\,{\sc i}}\,}
\newcommand{\HIbf}{\mbox{H\hspace{0.155 em}{\footnotesize \bf I}}}
\newcommand{\HIit}{\mbox{H\hspace{0.155 em}{\footnotesize \it I}}}
\newcommand{\HIsl}{\mbox{H\hspace{0.155 em}{\footnotesize \sl I}}}
\newcommand{\HIss}{\mbox{H\,{\sc i}}}
\newcommand{\HII}{\mbox{H\,{\sc ii}}}
\newcommand{\kms}{\mbox{\rm km\,s$^{-1}$}}
\newcommand{\kmsMpc}{\mbox{\rm km\,s$^{-1}$\,Mpc$^{-1}$}}
\newcommand{\LB}{\mbox{$L_{B}$}}
\newcommand{\LBnul}{\mbox{$L_{B}^0$}}
\newcommand{\LBsun}{\mbox{$L_{\odot,B}$}}
\newcommand{\Lsun}{\mbox{$L_{\odot}$}}
\newcommand{\LsunMsun}{\mbox{$L_{\odot}$/${\cal M}_{\odot}$}}
\newcommand{\LFIR}{\mbox{$L_{FIR}$}}
\newcommand{\LFIRLB}{\mbox{$L_{FIR}$/$L_{B}$}}
\newcommand{\LFIRLBnul}{\mbox{$L_{FIR}$/$L_{B}^0$}}
\newcommand{\LFIRLsun}{\mbox{$L_{FIR}$/$L_{\odot,Bol}$}}
\newcommand{\MHI}{\mbox{${M}_{\rm HI}$}}
\newcommand{\MHILB}{\mbox{${M}_{\rm HI}$/$L_{\rm B}$}}
\newcommand{\MHILBnul}{\mbox{${M}_{HI}$/$L_{B}^0$}}
\newcommand{\Msun}{\mbox{${\rm M}_\odot$}}
\newcommand{\MsunLsun}{\mbox{${\rm M}_{\odot}$/$L_{\odot,Bol}$}}
\newcommand{\MsunLBsun}{\mbox{${\rm M}_{\odot}$/$L_{\odot,{\rm B}}$}}
\newcommand{\MT}{\mbox{${\cal M}_{\rm T}$}}
\newcommand{\MTLBnul}{\mbox{${\cal M}_{T}$/$L_{B}^0$}}
\newcommand{\MTLBsun}{\mbox{${\cal M}_{T}$/$L_{\odot,B}$}}
\newcommand{\mi}{\mbox{$\mu$m}}
\newcommand{\nan}{\mbox{Nan\c{c}ay}}
\newcommand{\NH}{\mbox{N$_{HI}$}}
\newcommand{\OIII}{\mbox{[O\,{\sc iii}]}}
\newcommand{\s}{\mbox{$\sigma$}}
\newcommand{\Tb}{\mbox{$T_{b}$}}
\newcommand{\tis}[2]{$#1^{s}\,\hspace{-1.7mm}.\hspace{.1mm}#2$}
\newcommand{\vrot}{\mbox{$v_{rot}$}}

\def\sun{\hbox{$\odot$}}
\def\earth{\hbox{$\oplus$}}
\def\la{\mathrel{\hbox{\rlap{\hbox{\lower4pt\hbox{$\sim$}}}\hbox{$<$}}}}
\def\ga{\mathrel{\hbox{\rlap{\hbox{\lower4pt\hbox{$\sim$}}}\hbox{$>$}}}}
\def\sq{\hbox{\rlap{$\sqcap$}$\sqcup$}}
\def\arcmin{\hbox{$^\prime$}}
\def\arcsec{\hbox{$^{\prime\prime}$}}
\def\fd{\hbox{$.\!\!^{d}$}}
\def\fh{\hbox{$.\!\!^{h}$}}
\def\fm{\hbox{$.\!\!^{m}$}}
\def\fs{\hbox{$.\!\!^{s}$}}
\def\fdg{\hbox{$.\!\!^\circ$}}
\def\farcm{\hbox{$.\mkern-4mu^\prime$}}
\def\farcs{\hbox{$.\!\!^{\prime\prime}$}}
\def\fp{\hbox{$.\!\!^{\scriptscriptstyle\rm p}$}}
\newcommand{\etal}{{et~al.}\,}     
\newcommand{\eg}{{e.g.},\ }         
\newcommand{\ie}{{i.e.},\ }         
\newcommand{\cf}{{cf.},\ }          

\thesaurus{03(11.04.1;   
              11.07.1;   
              11.09.4;   
              13.19.1)}  

\title {Nan\c{c}ay ``blind'' 21cm line survey of the Canes Venatici
group region}

\author{R.C.~Kraan-Korteweg\,\inst{1,2}, W.~van Driel\,\inst{3},
        F.~Briggs\,\inst{4}, B.~Binggeli\,\inst{5}, 
        \and T.I.~Mostefaoui\,\inst{2}}

\offprints{R.C.~Kraan-Korteweg, Universidad de Guanajuato, Mexico; 
e-mail : kraan@astro.ugto.mx \rm}   

\institute{Departemento de Astronomia, Universidad de Guanajuato, Apartado Postal 144
 Guanajuato, GTO 36000, Mexico
\and
DAEC, Observatoire de Paris, 5 Place Jules Janssen, 
 F-92195 Meudon Cedex, France
\and
Unit\'e Scientifique \nan, CNRS USR B704, Observatoire de Paris, 
 5 Place Jules Janssen, F-92195 Meudon Cedex, France
\and
Kapteyn Astronomical Institute, Groningen University, P.O. Box 800, 
 NL-9700 AV Groningen
\and
Astronomical Institute, University of Basel, Venusstrasse 7,
 CH-4102 Binningen, Switzerland} 

\date{{\it Final version}: 29 Octobre 1998}
\maketitle

\markboth{Kraan-Korteweg \etal:  A blind H\,{\footnotesize I} search
in the CVn Group Region}{xxx}

\begin{abstract}  

A radio spectroscopic driftscan survey in the 21cm line
with the \nan\ Radio Telescope of 0.08 steradians of sky
in the direction of the constellation Canes Venatici covering
a heliocentric velocity range of  $-350 < V_{hel} < 2350$~\kms\
produced 53 spectral features, which was further reduced to a sample
of 33 reliably detected galaxies by extensive follow-up observations.
With a typical noise level of rms = 10 mJy after Hanning smoothing,
the survey is -- depending on where the detections are located with 
regard to the center of the beam -- sensitive to 
$M_{HI} = 1-2 {\times}10^8 h^{-2}$\Msun\ 
at 23 $h^{-1}$Mpc and to 
$M_{HI} = 4-8 {\times}10^7 h^{-2}$\Msun\ 
throughout the CVn Groups.

The survey region had been previously examined on deep optical 
plates by Binggeli \etal (1990) and contains loose groups with many 
gas-rich galaxies as well as voids. No galaxies that had not 
been previously identified in these deep optical surveys were 
uncovered in our \HI survey, neither in the groups nor the voids.  
The implication is that 
no substantial quantity of neutral hydrogen contained in 
gas-rich galaxies has been missed in these well-studied groups.
All late-type members of our sample are listed in the Fisher
and Tully (1981b) optically selected sample of nearby late-type
galaxies; the only system not contained
in Fisher and Tullys' Catalog is the S0 galaxy NGC 4203.
Within the well--sampled CVn group volume with distances corrected
for flow motions, the \HI\ mass function is best fitted with the Zwaan 
\etal (1997) \HI\ mass function  ($\alpha=-1.2$)
scaled by a factor of f=4.5 in account of the locally overdense region.
\end {abstract}


\section{Introduction} 
The proposed existence of a possibly very large population 
of ``low surface brightness'' 
(LSB) galaxies has implications for the mass density of the Universe,
galaxy evolution, testing large-scale structure theories, and the 
interpretation of high-redshift quasar absorption-line systems. 
The evidence for the existence of a considerable population of 
such LSB galaxies is increasing; they are so far, however, mainly detected 
at larger redshifts. Recent LSB galaxy surveys seem to be strongly
biased against the detection of nearby systems.

Deep optical surveys using new POSS-II plate material (Schombert 
\etal 1997), automated plate scanners (Sprayberry \etal 1996), and 
clocked-CCD drift-scan surveys (Dalcanton \etal 1997) are succeeding 
at identifying faint LSB galaxies. Since many of these are
gas-rich, they can be detected with sensitive radio observations 
(Schombert \etal 1997, Sprayberry \etal 1996 and Impey \etal 1996, 
Zwaan \etal 1998). 

One concern is how to
appraise the mass density (both for optically luminous material
and for neutral gas) contained in these newly identified populations
in comparison to what has historically been included using the older catalogs
that were based on higher surface brightness selection.  Briggs (1997a)
addressed this problem and noted that the new LSB catalogs contain
galaxies at systematically greater distances than the older catalogs,
and estimates of luminosity functions, \HI-mass functions and integral
mass densities are not substantially altered by the addition of
these new catalogs. Either the new surveys are not sensitive to nearby, large
angular diameter LSB galaxies, or the nearby LSB disks were already
included in the historical catalogs because their inner regions are
sufficiently bright that at small distances they surpass the 
angular diameter threshold to make it into catalogs.  

The question still
remains whether a population of nearby ultra-low surface brightness
objects with \HI\ to optical luminosity ratios \MHILB\ far in excess
of 1 could have escaped identification even to this day. For this
reason we have performed a systematic, sensitive (rms = 10 mJy
after Hanning smoothing) blind \HI-line survey in driftscan mode in 
the nearby universe  ($-350 < V_{hel} < 2350$~\kms)
using the \nan\ Radio Telescope. A blind \HI-survey
will directly distinguish nearby gas-rich dwarfs at low redshifts
from intrinsically brighter background galaxies and allow us
to study the faint end of the \HI\ mass function. 

In order to directly compare the mass density detected in neutral 
hydrogen with the optically luminous mass density we have selected the
Canes Venatici region (CVn), a region which contains both loose galaxy 
groups and voids and which had previously been surveyed for 
low-luminosity dwarfs on deep optical plates (Binggeli \etal 1990, 
henceforth BTS).

The \nan\ CVn Group blind survey scanned
this volume of space at a slightly deeper \HI\ detection level than the
pointed observations that Fisher and Tully (1981b) made of their
large optically selected sample.  Thus, these new observations 
allow at the same time a test of the completeness of late-type
galaxies in this historical catalog.

In Section 2, the region observed with the \nan\ telescope is
described, followed by a description of the telescope, the driftscan
observing mode as well as the reduction and data analysis procedures.
In Section 3, the results from the driftscan survey are given
including a detailed discussion on the sensitivy and completeness
limit of the survey (3.1), the results from pointed follow-up 
observations of galaxy candidates without an unambigous optical 
counterpart on the one hand (3.2.2), and deep pointed observations of 
optically identified dwarfs not detected in the blind \HI\ survey
on the other hand (3.2.3). This is followed by a comparison of the
driftscan data with pointed observations (3.3). 
In section 4, the global properties of the \HI-detected galaxies
are discussed including the effects of Virgocentric flow on the
derived parameters and a discussion of the discrepancy between the
observed velocity and the independent distance determination of the
galaxy UGC~7131 (4.1). This is followed by a discussion of the
depth of this survey in comparison to the Fisher--Tully (1981b) Nearby Galaxy
Catalog (4.2) and the \HI-mass function for the here performed blind
\HI\ driftscan survey (4.3). In section 5, the conclusions are summarized.

\section{Observations} 

\subsection{The observed region} 

The region of our blind \HI\ survey is a strip bounded in declination
by  $+29\dgr08\arcmin < \delta < +35\dgr22\arcmin$ and runnning in R.A. 
from 11$^h30^m$ to 15$^h00^m$. The choice of this area
was motivated primarily by the pre-existence of the deep optical survey of
Binggeli \etal 1990 (BTS), which covers roughly half of our
region -- essentially all between 12$^h$ and 13$^h$ plus the southern half of
our strip, with a gap between 13$^h20^m$ and 13$^h45^m$ (\cf Fig.~3 of BTS). 
This sky region was regarded as ideal because it captures 
a big portion of the nearby Canes Venatici cloud of galaxies -- a prolate
structure seen pole-on ($V_{hel} \sim$ 200 - 800~\kms) running across our strip 
between $12^h$ and $13^h$ (\cf Fig.~\ref{cone.fig} and
Fig.~\ref{supergal.fig}). Superposed on this cloud, but shifted to the South, 
almost out of our survey strip, lies the somewhat more distant Coma I cluster 
at $<V> \sim$ 1000~\kms\ (\cf de Vaucouleurs 1975, Tully \& Fisher
1987, and BTS). 
Aside from these two galaxy aggregates, there are only a handful of field 
galaxies known in the area out to $V \le $ 2000~\kms\ (BTS, Fig.~4). 
Hence, we have on the one hand a very nearby, loose group or cloud (CVn) 
with many known faint, gas-rich dwarf galaxies with which we can
crosscorrelate our blind detections. On the other hand, we have a void region 
where the expectation of detecting previously unknown, optically
extremely faint \HI-rich dwarfs could be large.

The optical survey of BTS was based on a set of long (typically 2 hour) 
exposure, fine-grain IIIaJ emulsion Palomar Schmidt plates which were 
systematically inspected for
all dwarf galaxy-like images. A total of 32 ``dwarfs'' and dwarf candidates 
were found in the region of our blind \HI-survey strip, 23 of which were
judged to be ``members'' and possible members of the CVn cloud. The
morphological types of these objects are almost equally distributed between 
very late-type spirals (Sd-Sm), magellanic irregulars (Im), early-type dwarfs 
(dE or dS0), and ambiguous types (dE or Im; for dwarf morphology see Sandage
\& Binggeli 1984). Not included here are bright, normal galaxies 
like Sc's, which are certainly rediscovered by the \HI-survey.

The ``depth'' of the BTS survey is given through the requirement that 
the angular 
diameter of a galaxy image at a surface brightness level of 25.5 
Bmag arcsec$^{-2}$ had to be larger than $0\farcm2$. This completeness 
limit turned out to be close to,
but not identical with, a total apparent blue magnitude $B_T$ = $18^m$
(\cf Fig.~1 of BTS). For galaxies closer than about 10 $h^{-1}$ Mpc 
(or $V_{hel} \sim$ 1000~\kms), \ie the distance range of the CVn cloud,
in principle all galaxies brighter than $M_{\rm B_{\rm T}}$ = $-12^m$ 
should be catalogued; out to 23 $h^{-1}$ Mpc (or $V_{hel} \sim$
2300~\kms), the volume limit of our survey, this value is close 
to $M_{\rm B_{\rm T}}= -14^m$. It should be 
mentioned that these limits are valid only for ``normal'' dwarf galaxies which 
obey the average surface brightness-luminosity relation (the ``main sequence''
of dwarfs, \cf Ferguson \& Binggeli 1994). Compact dwarfs, like BCDs and 
M32-type systems, can go undetected at brighter magnitudes, though they seem 
to be rare in the field at any rate.

What \HI\ detection rate could be expected for such an optically selected
dwarf sample of the mentioned optical depth? Our blind \HI\ survey was 
designed to be maximally sensitive for nearby ($V < $ 1000~\kms), low HI-mass
dwarf galaxies with \HI-masses of a few times $10^7$ \Msun. Assuming 
the complete sample of 
dwarf irregular galaxies in the Virgo cluster outskirts as representative, 
an \HI\ mass of $10^7$ \Msun\ corresponds -- on average -- to a total 
blue luminosity, $\LB$, of $2$ x $10^7$ \LBsun\ (Hoffman \etal 1988), which 
in turn roughly corresponds to $M_{\rm B_{\rm T}}$ = -13$^m$. 
Based on the optically defined completeness limit, the depth of our 
blind \HI\ survey is comparable to the optical depth of the BTS survey
for ``normal'' gas-rich dwarfs.
Thus our blind survey would uncover only new galaxies if the faint end
of the \HI\ mass function is rising steeply and/or if there exists
a population of LSB dwarf galaxies with high \MHILB\ ratios which
fall below the optical completeness limit of the CVn survey.

As it turned out (see below), and in contradiction to what
was found in a similar survey in the CenA group region (Banks \etal 1998), no 
extra population of extremely low-surface brightness (quasi invisible) 
but \HI-rich dwarf galaxies was found in the volume surveyed here.

\subsection{The \nan\ Radio Telescope} 

The \nan\ decimetric radio telescope is a meridian transit-type 
instrument with an effective collecting area of roughly 7000~m$^{2}$ 
(equivalent to a 94-m parabolic dish). Its unusual configuration 
consists of a flat mirror 300 m long and 40 m high, which can be tilted
around a horizontal axis towards the targeted declination, and a fixed spherical 
mirror (300 m long and 35 m high) which focuses the radio waves towards
a carriage, movable along a 90 m long rail track, which houses the receiver 
horns. Due to the elongated geometry of the telescope  
it has at 21-cm wavelength a half-power beam width of \am{3}{6} 
$\times$ 22$'$ ($\alpha$ $\times$ $\delta$). 
Tracking is generally limited to about one 
hour per source for pointed observations. Typical system temperatures are  
$\sim$50~K for the range of declinations covered in this work.

\subsection{Observational stragegy} 

The observations were made in a driftscan mode in which the
sky was surveyed in strips of constant declination over the
R.A. range 11$^h30^m$--15$^h00^m$. 

During each observation, the flat telescope mirror was first tilted towards
the target declination and the focal carriage was then blocked in place on its
rail, near the middle of the track where the illumination of the
mirror is optimal; tests in windy conditions showed that its position 
remained stable to within a millimeter or so, quite satisfactory for 
our purpose.

The procedure required fixing the telescope pointing coordinates each day 
to point at a chosen starting coordinate in the sky as tabulated in
Table 1. These coordinates for the start of the scan
are given for the 1950 equinox. But since the observations were
made over the period of January 1996 to Novembre 1996, the observed 
survey strips lie along lines of constant declination of epoch
1996.5. In order to accumulate
integration time and increase the sensitivity along the survey
strips, each strip was scanned 5 times on different days.

\input table1.tex

We adopted a search strategy with a full 4$'$ beamwidth sampling rate
in R.A. (integrating 16 seconds per spectrum, followed by an 0.3s second
read-out period).  The 1024 channel autocorrelation spectrometer was
divided to cover two slightly overlapping 6.25 MHz bands 
centered on $V=324$~\kms\  and $V=1576$~\kms\ in two
polarizations. Since each declination strip was traced 5 times, and there
were 17 declination strips observed (spaced by the full 22$'$ beamwidth
in declination), a total of more than 250,000 individual spectra were
recorded in the course of the survey phase of this project.  
A complete strip obtained on one day consisted of 40 ``cycles'' with
each cycle comprising 20 integrations of 16 seconds.  In many instances,
the telescope scheduling prevented us from obtaining the full number
of cycles, as is summarized in the Table 1, where the survey's
sky coverage is listed with  `Dec' indicating the center
declination of each strip and the number of cycles obtained on each
day is listed.

\subsection{Data reduction and analysis} 

The logistical problem of calibrating, averaging and displaying this
large quantity of data was simplified in the following way:  

\noindent (1) The
raw spectra were converted from the telescope format to binary
compatible files for processing in the ANALYZ package (written 
and used at the Arecibo Observatory). 

\noindent (2) Using ANALYZ and IRAF, the data from
each day's driftscan were formatted into an IRAF image format, one
image per day, with successive spectra filling successive
lines in the image. In this way, each of the $17{\times}5=85$ 
driftscans in Table 1 could be viewed and manipulated as a single
data block. The technique had been used earlier in the Arecibo survey
reported by Sorar (1994) (see also Briggs \etal 1997, Zwaan \etal 1997).

\noindent (3) The first important processing
step is calibration of the spectral passband.
This was accomplished by averaging all the spectra from
a single day's driftscan (after editing
spectra corrupted by radio interference, strong \HI\ emission from
discrete galaxies, or instrumental problems) into one high signal-to-noise
ratio spectrum.  Each line of the data image was then divided by this 
``passband''.

\noindent (4) The 5 passband-calibrated data images from each declination
strip were blinked against each other in order to verify system 
stability. Since continuum sources appear as lines in these images, the
timing and relative strength of these lines confirm the timing of the
integrations throughout the driftscan and the gain stability of the
receivers. Examples of these images are shown by Briggs \etal (1997,
Fig.~1).

\noindent (5) Continuum substraction was performed by fitting a quadratic
baseline separately 
to each line of the data image, after masking out the portion
of the spectrum containing the local Galactic emission.  Further editing
was performed as necessary after inspection of the continuum-subtracted
images.

\noindent (6) The data images for each declination were averaged, the
spectral overlap removed, the two polarizations averaged, and the data
Hanning smoothed in the spectral dimension to remove spectral ringing.
This resulted in a velocity resolution of 10~\kms\ and an rms of
typically 10 mJy.

\noindent (7) At this stage, the calibrated data could be inspected and
plotted in a variety of ways:  Simple image display and optical 
identification of galaxy candidates (accompanied by full use of the
image display tools in IRAF and GIPSY for smoothing and noise analysis),
plots of individual spectra with flux density as a function of
frequency, or display as a three dimensional cube of R.A., Dec. and velocity.

Visual image inspection is highly effective at finding galaxy candidates, 
particularly in uncovering extended (in postion as in velocity spread)
low-signal features.  But
vigorous application of a statistical significance criterion is necessary
to avoid selection of tempting but not statistically significant
candidates. In this survey, the threshold was set at $4~\sigma$, which
led to a large number of non-confirmed features once the follow-up 
observations were made (\cf section 3.2.2).

\section{Results} 
With the adopted blind survey strategy and the stability of the 
telescope system we were able to consistently obtain the low 
rms noise levels we expected to reach:
25 mJy at full spatial and velocity resolution, and 10 mJy after 
Hanning smoothing.

\begin{figure*} 
\hfil\epsfxsize 18cm \epsfbox{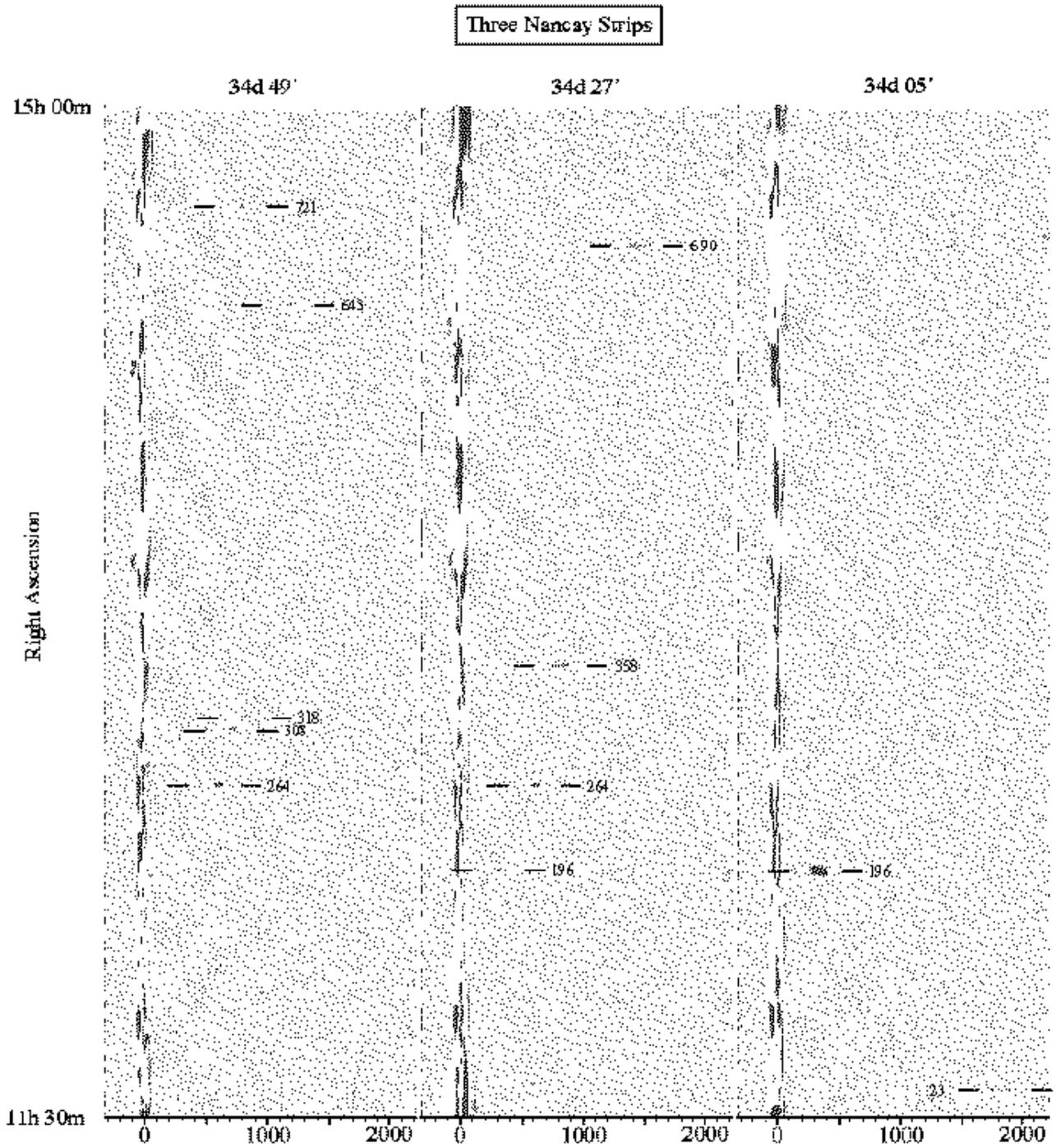}\hfil
 \caption[]{A `clean' spectrum, Hanning-smoothed in RA and velocity
(rms = 10 mJy), of 3 adjacent strips of the \HI\ drift-scan survey in the CVn
constellation. The figure contains about 48 000 individual
spectra. Eleven extragalactic signals from 9 galaxies are visible in
these strips.}
  \label{strips.fig}
 \end{figure*}

As an example of typical results, Fig.~\ref{strips.fig} displays 
the images (``cleaned'' and Hanning-smoothed in velocity and
R.A.) for 3 adjacent declination strips of our survey
The velocity range is indicated on the horizontal axis
and the R.A. range (800 spectra) on the vertical axis. 
The strong positive and negative signals around 0~\kms\ are due to 
residuals of the Galactic \HI\ emission. Some standing waves 
due to this emission entering the receiver are visible in the lower 
velocity band.

%

On the three images displayed here, eleven extragalactic signals from 
9 galaxies can be identified. We inspected the images of the 17 strips 
of the survey carefully by eye and listed all candidates 
as well as all detections above the 4~$\sigma$ threshold.

The so identified galaxy candidates were run through the NED
and LEDA databases for crossidentifications. The match in
positions, velocity if available, morphology and orientation were
taken into account. 

The positions of detections identified in the driftscan 
survey were determined according to the line number: sequence in R.A.
of the 800 contiguous spectra (\cf the 11 detections 
in Fig.~\ref{strips.fig}) as R.A. = $11^h30^m + (16\fs33 \times line + 8^s$).
The 16 seconds reflect the integration time per spectrum
plus the 0.33 second per read-out period. The positions (R.A. and Dec)
were interpolated if the signal appears in various lines or 
strips according to the strength of the signal. The positional accuracy
turned out to be remarkably precise in R.A. over the whole R.A. range
(\cf Table~2a) and reasonable in Dec if a candidate was evident
in more than one strip.

If no clear-cut crossidentification could be made, the POSS I and II 
sky surveys were inspected. If 
this did not yield a likely counterpart, the five scans leading to
the final images were inspected individually to decide whether the
signal was produced by radio interference in an individual scan.  

In this manner, 53 galaxy candidates were retained from the 17
\nan\ strips. Of these galaxy candidates, 33 could be identified
with an optical counterpart, 30 of which have published \HI\ 
velocities in agreement with our driftscan measurements.

For the 20 galaxy candidates without a clear optical counterpart,
pointed follow-up observations were made with the \nan\ telescope.
These follow-up observations are described in Section 3.2.2. and allow the 
establishment of a database of blind \HI\ detections in the 
volume sampled, which is reliable down to the 4~$\sigma$ level.
Interestingly, none of the candidates could be
confirmed. But, note that for the number of independent measurements 
that were obtained in this survey at, say 50~\kms\ resolution, 
one expects to obtain of the order of 15 positive and 15 negative
deviations exceeding $4~\sigma$ purely by chance.  

\subsection{Sensitivity of the survey} 

The sensitivity of the survey varies with position in the sky due
to the adopted strategy, which was a result of logistical concerns of
data storage during the driftscan. A system temperature of 50K was 
adopted for this declination range.

The observing technique recorded the integrations
every 16.3 seconds, during which time the beam drifted through the
sky by a full Half-Power-Beam-Width. In comparison,
a ``pointed'' integration, with the
source of interest located on the beam axis, would reach a 
sensitivity of about 
$\sigma_{rms} \approx 0.025(\Delta V/5~\kms)^{-1/2}$ Jy for
velocity resolution $\Delta V$ after 5~$\times$~16.3 seconds
integration with dual
polarization. This would lead to a $4~\sigma$ detection limit for
integral line fluxes $I \approx 1.4(\Delta V/30)^{1/2}$~\kms.
However, sources that pass through the beam do not appear in the
data at the full strength they have when observed at the
peak of the beam for the full integration time. Instead, a source
that traverses the HPBW from one half-power point to the other
suffers a loss in signal-to-noise ratio of a factor of 0.81,
and a source that chances to pass the center of the beam at the boundary
of two integrations suffers a factor of 0.74.  (Had samples been taken
at 8 second intervals, these factors would be 0.84 and 0.80.)

An additional loss of S/N occurs because the survey declination strips
were spaced by a full beamwidth (22$'$). A source falling 11$'$ from the
beam center is observed at half power. Some of this loss in sensitivity
is recovered by averaging adjacent strips, so that these sources
midway between the two strip centers then experience a net loss in S/N
of $1/\sqrt{2}$ relative to the strip center.

\begin{figure}
\psfig{figure=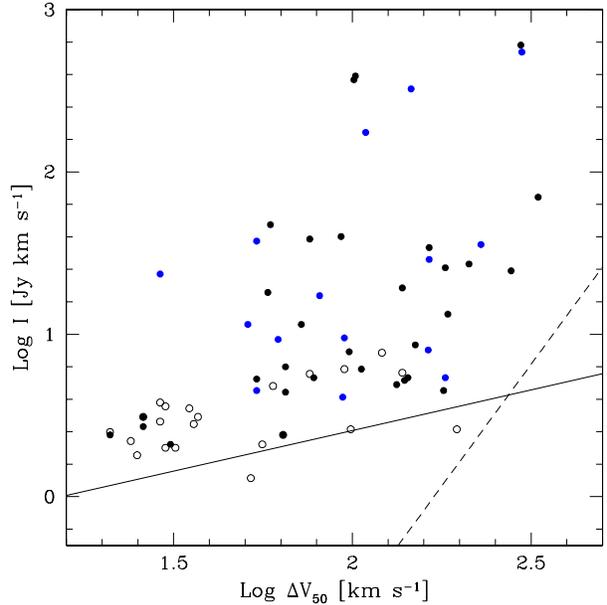,width=8.5cm,angle=0}
  \caption[]{\HI\ line flux for each galaxy plotted as a function
of linewidth measured at the 50\% level.  Filled dots are confirmed
detections; unfilled circles are false detections that were not
confirmed in the ``pointed'' follow-up observations. The solid
line indicates the $4~\sigma$ detection level.
Detections are not expected below
the diagonal dashed line at the lower right.}
  \label{sensit.fig}
\end{figure}

Combining the loss factors incurred due to R.A. sampling and declination
coverage shows that the S/N is reduced relative to pointed observations
by factors ranging from 0.81 to ${\sim}0.5$.  Thus there is a range 
to the $4~\sigma$ detection limit.
It is clear from the distribution of non-confirmable signals 
(\cf Fig.~\ref{sensit.fig})
that there is also non-Gaussian noise at work, such as that resulting
from radio interference, leading to a few spurious signals with 
apparent significance of as much as ${\sim}10~\sigma$.

The \HI\ line flux and the profile velocity width are two 
measured quantities.  Fig.~\ref{sensit.fig} shows the integral
line flux for each galaxy plotted as a function of its velocity width.
The detection limit (delineated in Fig.~\ref{sensit.fig}) rises as 
$I_{det}{\propto}\Delta V^{1/2}$, since the signal $I$ from a 
galaxy with flux density $S_{\nu}$ grows as $I=S_{\nu}\Delta V$, 
while the noise $\sigma$ in measuring a flux density is distributed 
over $\Delta V$ is $\sigma \propto (\Delta V)^{-1/2}$, so
that  the signal-to-noise ratio $S_{\nu}/\sigma \propto (\Delta V)^{-1/2}$
drops as a signal of constant $I$ is spread over increasing width.
The detection boundary would be a band for true fluxes due to the 
lower survey sensitivity for detections offset from the center of 
the survey strip. 
The dashed line in the figure marks a region to the lower 
right that is not
expected to be populated, since galaxies of large velocity width typically
have large \HI\ masses, and, to have such a low measured flux, these galaxies
would lie beyond the end of the survey volume, which is limited by the
bandwidth of the spectrometer.

\subsection{The Data} 
\subsubsection{Confirmed Detections} 
The 33 reliable detections are listed in Table 2a. The 
left-hand part of the table lists the \HI\ parameters as derived from
the driftscan observations. The first column gives the running number
of all our 53 galaxy candidates ordered in R.A. The second column
gives the the R.A. as explained in section 3 
(R.A. = $11^h30^m + (16\fs33 \times line) + 8^s$) and the third column
the central declination of the strip on which the signal was detected.
Even with the large offset from one declination strip to the
next of the \nan\ radio telescope (22$\arcmin$ HPBW),
strong sources were, amazingly enough, visible on up to
four adjacent strips. 

\input table2a.tex

Within a declination strip, the \HI\ parameters were measured 
from the sum of the lines in which a candidate was detected 
using the Spectral Line Analysis Package by Staveley-Smith (1985).
We list here the heliocentric central line velocity using 
the optical convention ($V=\frac{c(\lambda - \lambda_{o})}{\lambda_{o}}$), 
$V_{\rm HI}$, the  profile FWHM, $W_{50}$, the profile width measured at 20\% 
of the peak flux density level, $W_{20}$, the integrated line flux, I, and the 
logarithm of the \HI\ mass, $M_{\rm HI}$. The latter should be
regarded as preliminary as the \HI\ mass was determined
for the strongest appearance of the signal and is uncorrected
for the reduction in observed flux (described in section 3.1.) due 
to offset from the center of the beam. Moreover, 
the adopted distances were taken straight from the heliocentric 
\HI\ velocity $V_{hel}^{\rm HI}$ using a Hubble constant H$_0$=100~\kmsMpc.

For each \HI\ detection a possible optical identification is given.
Optical redshifts are known for 17 optical candidates, though 30 do
have independent \HI\ velocities (\cf Table~4). The listed optical data 
are mainly from the NED and LEDA databases; $D$ and $d$ are the optical major and 
minor axis diameters, respectively, and $V_{\rm opt}$ is the
heliocentric optical systemic velocity.

\subsubsection{Pointed observations of galaxy candidates} 

As noted under results, 20 likely detections were identified above the
$\approx 4~\sigma$ level for which no credible optical counterpart
could be identified either in the NED and LEDA databases or by visual
examination on the sky surveys. These possible candidates generally are weak
`signals' and in order to check on their reality we obtained pointed 
follow-up observations with the \nan\ telescope.
These follow-up observations allow the establishment of a database 
of blind \HI\ detections in the volume sampled here which is reliable 
down to the 4~$\sigma$ level. The galaxy candidates are listed in Table~2b
with the equivalent parameters as given in Table~2a. Also listed, 
next to the integrated flux, is the factor above the 1~$\sigma$ noise
level to compare it with our 4~$\sigma$ detection limit 
$I \approx 1.4(\Delta V/30)^{1/2}$~km~s$^{-1}$ defined in section 3.1. 

\input table2b.tex

These follow-up's resulted in an rms noise of about 6 mJy on
average  (compared to 10 mJy for the survey). 
We obtained our observations in total power (position-switching) mode using 
consecutive pairs of two-minute on- and two-minute 
off-source integrations. The autocorrelator was divided into two pairs
of cross-polarized (H and V) receiver banks, each with 512 channels 
and a 6.4~MHz 
bandpass. This yielded a velocity resolution of $\sim$5.2~\kms, which was
smoothed to 10.5~\kms, when required, during the data analysis. 
The center frequencies of the two banks were tuned to the radial
velocity of the galaxy candidate (\cf Table~2b).

The follow-up observations first were repeated on the optimized
position from the driftscan detection. If not confirmed, a second round
of pointed observations was made, offset by half a beam (11$\arcmin$)
to the north and south of the original position while the original
R.A. position which can be determined to higher accuracy was kept
constant. In a few cases, a further search at quarter beam offsets
was done as well. The follow-up observations consisted of a number of
cycles of 2 minutes on- and 2 minutes off-source integration. 
Off-source integrations were taken at 
approximately 25$\arcmin$ east of the target position.

We reduced our follow-up pointed \HI\ spectra using 
the standard \nan\ spectral line reduction 
packages available at the \nan\ site. With this software 
we subtracted baselines (generally third order polynomials), averaged 
the two receiver polarizations, and applied a declination-dependent
conversion factor to convert from  units of T$_{sys}$ to flux density 
in mJy.  The T$_{\rm sys}$-to-mJy conversion factor is determined via 
a standard 
calibration relation established by the \nan\ staff
through regular monitoring of strong continuum sources. This 
procedure yields a calibration accuracy of $\sim\pm$15\%.
In addition, we applied a flux scaling factor of 1.25 to our spectra based on 
a statistical comparison (Matthews \etal 1998) of recent \nan\ data with 
past observations. The derivation of this scaling factor 
was necessitated by the {\it a posteriori} discovery that the line 
calibration sources monitored at \nan\ by other groups 
as a calibration normalization check (see Theureau \etal 1997)
were very extended compared with the telescope beam, and hence would be 
subject to large flux uncertainties.

None of the candiates were confirmed, despite the fact that some of 
them have a rather large mean \HI\ flux density. Only 4 `candidates' 
are below our threshold. But, as mentioned above, for the number of 
independent measurements that were obtained in this survey 
we expect of the order of 15 positive and 15 negative
deviations exceeding $4~\sigma$ purely by chance. Many of the ``false''
detections have very low linewidths (around 30~\kms) and could be 
still be due to radio interference. 

The most disappointing candidate was No.~50. This object was
identified on two adajcent declination strips and on 2-3 adjacent
spectra in R.A. with very consistent \HI\ properties and a similar
high S/N ratio (8 -- 10) on both declination strips. It therefore seemed one of 
the most promising new gas-rich dwarf candidates in the CVn region.

\subsubsection{Optically known dwarf galaxies not detected in the
survey} 

Seven dwarf candidates of the BTS survey were not detected in the
blind survey. With the exception of BTS 160, these dwarfs were
observed individually to a lower sensitivity as the driftscan survey,
in the same manner as described in section 3.2.2.
We did not include BTS 160, because this dwarf had already been detected by
Hoffman \etal (1989) at Arecibo with 0.9 Jy~\kms. With an
average flux density of 26 mJy this dwarf was hence clearly below 
the threshold of our driftscan survey. Of the 6 remaining dwarfs, none 
were in fact detected, although the pointed observation of BTS 151
revealed a strong signal  with a flux of 20 Jy~\kms\ (\cf values in
brackets in Table~3), \ie a detection which should have popped up as 
about a 10~$\sigma$ detection in the driftscan mode. However, this 
signal matches exactly the velocity of the nearby
large spiral galaxy NGC 4656 (\cf Table~2a and 4, object 
No.~32) which is 45$\arcsec$ and 17$\arcmin$ away in R.A.
and Dec respectively, hence less than a beamwidth from the pointed observation.
The detected signal thus clearly originates from the spiral NGC 4656, and not
from the dwarf elliptical BTS 151.

\input table3.tex

The data of these observatons are summarized in Table~3.
Most of these possible dwarf members of the CVn group were 
not detected in \HI. This supports their classification as
early type dwarfs.

\subsection{Comparison of driftscan data with pointed observations}

Of the 33 reliable detections, 30 galaxies had \HI\ 
velocities published in the literature. For the 3 galaxies 
without prior \HI\ data we obtained follow-up observations
with the \nan\ radiotelescope.
We also obtained pointed follow-up observations for 8 galaxies already 
detected before in the \HI\ line, which seemed to merit an independent, 
new observation. The observations followed the procedures as described
in section 3.2.2.

The results from the driftscan images are summarized in Table~4
together with our new pointed observations and
\HI\ parameters from independent observations.
\input table4.tex

Overall, the agreement between the measurements obtained from the
driftscan survey and independent pointed observations are very
satisfactory with a few discrepancies discussed below:

\noindent
No.~12 (NGC 4173): The \HI\ line parameters of this 5$'$ diameter edge-on system
measured at Arecibo by Williams \& Rood (1986) and independently by
Bothun \etal (1985) are quite different from the other available
values.

\noindent
No.~13 (NGC 4203): This lenticular galaxy has an optical inclination of
about 20\dgr. Assuming that the gas rotates in the plane of the
stellar disk, very high \HI\ rotational velocities would be derived.
However, radio synthesis imaging (van Driel \etal 1988) has shown
that the outer gas rotates in a highly inclined ring at an inclination
of about 60\dgr, the value adopted in correcting the profile 
widths (see Table 5).

\noindent
For 4 galaxies (No.~33 = UGC 7916, No.~34 = UGCA 309, No.~35 = NGC
4861, No.~38 = CGCG 0999) the listed properties such as morphological 
type, diameters and magnitudes differ between the LEDA and NED.
The values in Table~2b and Table~5 are from LEDA.


We have compared the systemic velocities, line widths,
integrated line fluxes and \HI\ masses of the 33 reliable survey
detections with available pointed observations. It should be noted that the 
comparison data represent in no way a homogeneous set of measurements, as they
were made with various radio telescopes throughout the years; especially, care
should be taken that no \HI\ flux was missed in observations with the \am{3}{8}
round Arecibo beam.

A comparison of the difference between systemic velocities (actually,
the centre velocities of the line profiles) measured at \nan\ and 
elsewhere shows no significant dependence on radial 
velocity, with the exception of one data point for
NGC 4173 (No.~12). Here we consider the radial velocity of 1020~\kms\ 
measured at Arecibo by Williams \& Rood (1986) as spurious, seen 
the agreement between the 3 other measured values. The mean value 
and its standard 
deviation of the \nan--others systemic velocity difference
is 0.8$\pm$11.7~\kms\ (for a velocity resolution of 10.2~\kms\ at \nan). 

A comparison of the difference between the  $W_{50}$ and $W_{20}$ \HI\ line
widths measured at \nan\ and elsewhere shows a good correlation as well;
the largest discrepancy is found between the survey values for the
5$'$ diameter edge-on system NGC 4173 (No.~12) and the Arecibo
$W_{50}$ and $W_{20}$ measurements by Williams \& Rood (1986) 
as well as by Bothun \etal (1985), while 
the other pointed observations of this object are consistent with ours.

A comparison of the difference between the integrated \HI\ line
fluxes measured in the \nan\ survey and elsewhere shows 
a reasonable correlation. This indicates that the assumption of 
a constant (uncalibrated) system temperature of 50 K throughout the 
survey is justified.
3 larger discrepancies occur. This concerns No.~13 = NGC 4203 for which
the \nan\ survey flux is considerably higher than the flux measured at
Arecibo. As the galaxy's outer \HI\ ring is much larger than the Arecibo beam 
(see van Driel \etal 1988) the \nan\ measurement will reflect tht
total flux.  For No.~06 = BTS59 and No.~14 = UGC 7300, the \nan\ flux
is also considerably higher than a flux measured 
at Arecibo by Hoffman \etal (1989) , respectively Schneider \etal (1990); 
both are objects of about \am{1}{2} optical diameter only.

A comparison between the \HI\ masses derived straight from the
detections in the strip-images with pointed observations shows an
astonishingly narrow correlation from the lowest to the highest \HI\
masses detected in this survey. Overall, only a slight offset 
towards lower masses is noticeable for the driftscan results 
compared to pointed observations. This obsviously is due to the 
fact that the driftscan detects the galaxies at various offsets 
from the center of the beam.

\section{Discussion}

\subsection{Properties of the detected galaxies}
We derived a number of global properties for the 33 galaxies with 
reliable detections in our \HI\ line survey based on the available 
pointed \HI\ data, including our own new results. These are listed 
in Table~5. The values for the observed linewidths and the
heliocentric velocity are the means from all individual pointed
observations listed in Table~4.
The inclination, i, is derived from the cosine of the ratio of
the optical minor and major axis diameters, the profile widths $W_{50}$
and  $W_{20}$ were corrected to $W_{50}^{cor}$ and $W_{20}^{cor}$, 
respectively, using this inclination. The \HI-mass and blue absolute
magnitude, $M_{\rm B}$, were as a first approximation computed 
straight from the observed velocity and a Hubble constant of 
$H_o=100$ km~s$^{-1}$~Mpc$^{-1}$. We furthermore assumed that all
apparent magnitudes listed in Table 2 were measured 
in the B band.  

\input table5.tex

\begin{figure}
\psfig{figure=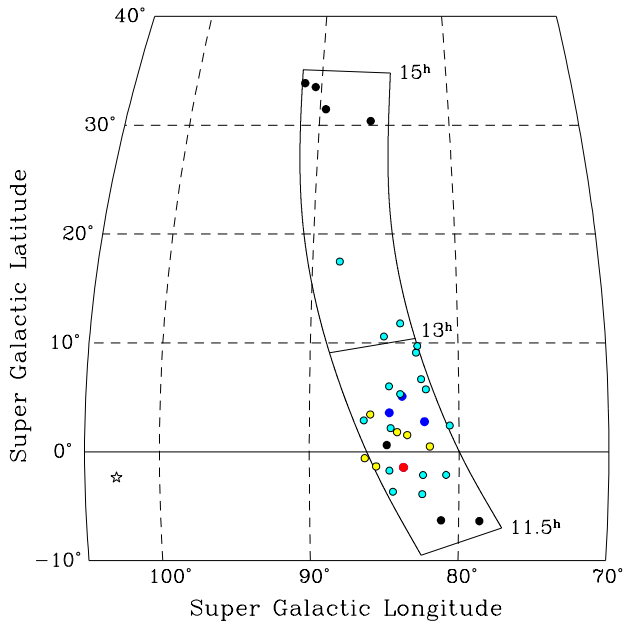,width=9.5cm,angle=0}
  \caption[]{Location of the detected galaxies in Supergalactic
coordinates. The survey area is bounded by solid lines; the
beginning R.A. at 11.5$^{h}$ and ending R.A. at 15$^{h}$ are
marked. The star symbol at the lower left indicates the position
of Virgo~A$=$M87.}
  \label{supergal.fig}
\end{figure}

The survey region lies quite close in the sky to the 
Virgo cluster. This is illustrated in Fig.~\ref{supergal.fig} which shows
the \nan\ survey region in Supergalactic coordinates, with the
positions of the in \HI\ detected galaxies as well as the apex of the 
Virgocentric flow field, M87. The proximity of our survey region
to the Virgo overdensity motivated us to calculate
distances to our galaxies using the POTENT program (Bertschinger \etal 1990),
hence corrected for Virgocentric infall. 

The POTENT corrections are based on the overall underlying
density field deduced from flow fields out to velocities
of 5000 \kms. A comparison of the correction with a pure Virgocentric
infall model (\cf Kraan-Korteweg 1986) confirms that the main
perturbation of the velocity field within our survey region is due
Virgocentric infall. More local density fluctuations or a Great
Attractor component (at an angular distance of $\approx$ 77\degr) 
have little impact on the velocities.

The ``absolute'' corrections to the observed velocity due the 
Virgo overdensity vary depending on velocity and angular distance 
from the Virgo cluster. But the effects on global properties such 
as magnitudes and luminosities and \HI-masses can be quite 
significant, particularly for low velocity galaxies at small 
angular distance from the apex of the streaming motion
(Kraan-Korteweg 1986).

The for streaming motions corrected velocities, absolute magnitudes 
and \HI-masses based on POTENT distances are also listed in Table~5.
The corrections in velocity reach values of nearly a factor of 2, 
the respective corrections in absolute magnitudes of $0\fm6$ 
and the logarithm of the \HI-masses of up to 0.6 dex.

\begin{figure}
\psfig{figure=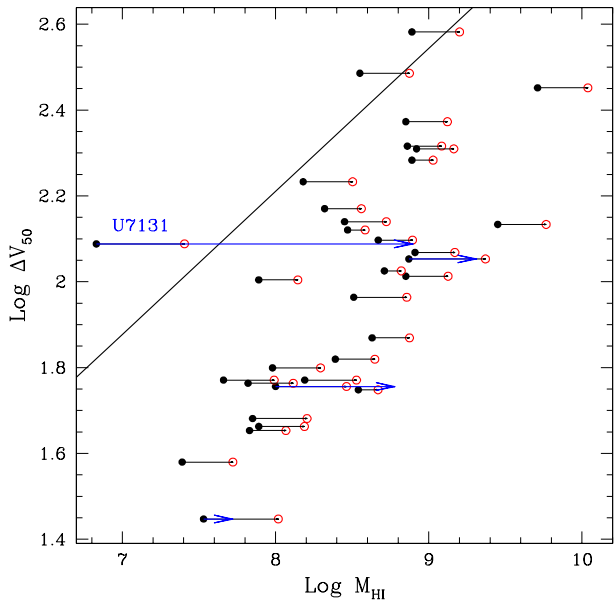,width=8.5cm,angle=0}
  \caption[]{Log of velocity width (50\%) as a function of \HI\ mass, 
$M_{HI}$, for the confirmed detections.
The line is a boundary to the \HI-mass versus line-width relation
derived from a much larger sample of galaxies (Briggs \& Rao 1993),
$\Delta V = 0.35 (M_{HI}/M_{\odot})^{1/3}$ km~s$^{-1}$.  \HI\ masses computed
using distance $d=v_{hel}/H_o$ are marked by solid dots. Open circles
mark masses corrected to POTENT distances to compensate for Virgocentric
flow, and the arrows indicate the masses for galaxies with independent
distance measures as referenced in the text.  
}
\label{vwid_MHI.fig}
\end{figure}

In Fig.~\ref{vwid_MHI.fig} the measured velocity width is plotted as 
a function of \HI\ mass.  There is a well known trend (a sort of 
\HI\ Tully-Fisher Relation) that larger $M_{HI}$ masses are strongly 
correlated with higher rotation speeds (\cf Briggs \& Rao 1993).  
Fig.~\ref{vwid_MHI.fig} displays \HI-masses based
on observed velocities as well as \HI-masses corrected for 
Virgocentric flow using the POTENT program 
(solid respectively open circles) including the shifts in galaxy
masses due to the perturbed velocity field.
The drawn line indicates an upper bound to the velocity width,
based on disk galaxies that are viewed edge-on; galaxies falling far 
below the line are viewed more face-on.   

A surprise that appeared in Fig.~\ref{vwid_MHI.fig} is that one
galaxy, UGC~7131, from our \nan\ survey lies above the usual bound 
for velocity width, even after correction for Virgocentric infall.

Subsequently, new measurements of distance using resolved stellar 
populations were released for four of our galaxies (Karachentsev 
\& Drozdovsky 1998, Marakova \etal 1998). This includes UGC~7131, 
which was found to lie at a distance $d>14$~Mpc, \ie considerably
further than indicated from the observed velocity listed in Table~5 
($V^{hel} = 251~\kms$) or for flow motions corrected velocity ($V^{POT} =
487~\kms$). However, with this new independent distance estimate
UGC~7131 does fall within the \HI-mass range expected for its linewidth.

Its morphology as evident
on the sky survey plates does not indicate a morphology earlier 
than the galaxy type listed in Table~2a of Sdm for which one could expect
a higher \HI-mass in agreement with its new determination (\cf shift in 
Fig.~\ref{vwid_MHI.fig}). It has a slight comet-like
structure not atypical for BCD galaxies. On the other hand, the deep CDD-image
in Markarova \etal (1998, their Fig.~3) finds UGC~7131 to be unresolved
and amorph, which does confirm the larger distance and is not
consistent with a nearby (low--velocity) galaxy.

Interestingly enough, the angular distance is a dominant parameter on the
infall pattern. UGC~7131 has a very small angular distance from the 
Virgo cluster, \ie only 19 degrees. If it were at a slightly smaller
angle, and depending on the model parameters for the virgocentric model, 
the solution for the distance would become triple valued: typically with one 
solution at low distance, one just in front of the Virgo cluster distance, and
one beyond the Virgo cluster distance (\cf Fig.~3 in Kraan-Korteweg
1986). Although the angular distance (from the Virgo cluster core)
within which we find
triple solutions does depend on the infall parameters such as the
decleration at the location of the Local Group, none of the models
with currently accepted flow field parameters suggests a triple
solution for galaxies with observed velocities as low as the one
measured for UGC~7131, except if the density profile within the Virgo
supercluster were considerably steeper than usually assumed.

\begin{figure}
\psfig{figure=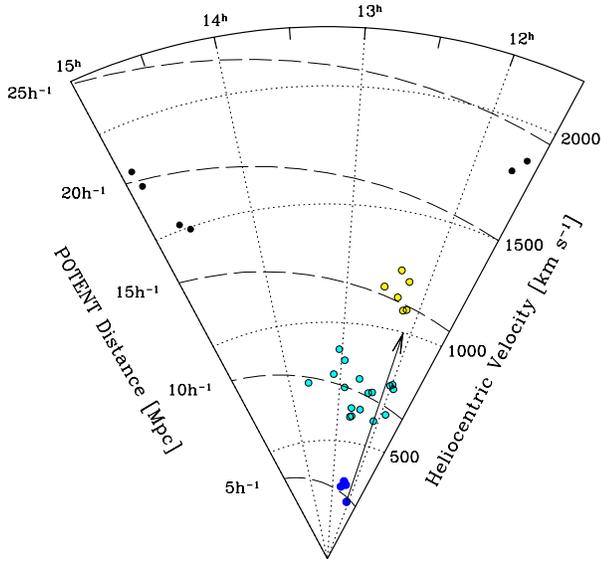,width=9.0cm,angle=0}
  \caption[]{Cone diagram showing the relative locations of
the detected galaxies as a function of R.A. and  
heliocentric redshift in km~s$^{-1}$. Long dashes show
contours of constant distance computed using POTENT (Bertschinger \etal 1990)
to compensate for Virgocentric flow. The arrow indicates
the revised distance for UGC~7131 (Karachentsev \& Drozdovsky 1998).
}
  \label{cone.fig}
\end{figure}

With the exception of the observed velocity, all further indications
about UGC~7131 support the considerably larger distance
--- even its distribution in redshift space.
The locations of the \HI\ selected galaxies are shown in a cone diagram
in Fig.~\ref{cone.fig} with the radial coordinate of Heliocentric
velocity and where the POTENT distances are drawn as contours.  
An arrow indicates the revision with regard to the location of
UGC~7131. It is clear from this display that UGC~7131 is not a member of
the nearby CVn I group, nor of the more distant
CVn II group, but most likely is a member of the Coma I group.
(Since our distances and survey volumes have been computed for convenience
using $H_o=100$ km~s$^{-1}$~Mpc$^{-1}$, the distances for these four
galaxies were adjusted to our scale, assuming that they were correct 
in a system with $H_o=75$ km~s$^{-1}$~Mpc$^{-1}$.)

Assuming that both the observed velocity and the revised distance to 
UGC~7131 are correct, this can only be combined if this galaxy resides
in a triple solution region of the Virgocentric flow pattern, implying
that our current knowledge of the density field within the Local 
Supercluster and the induced flow motions are not yet well established.
On the positive side, this example demonstrates that independent distance
derivations of fairly local galaxies, close in the sky
to the Virgo, can teach us considerably more about the density
field and the flow patterns within the Local Supercluster. 

\subsection{Comparison with the F-T  Catalog}
\begin{figure}
\psfig{figure=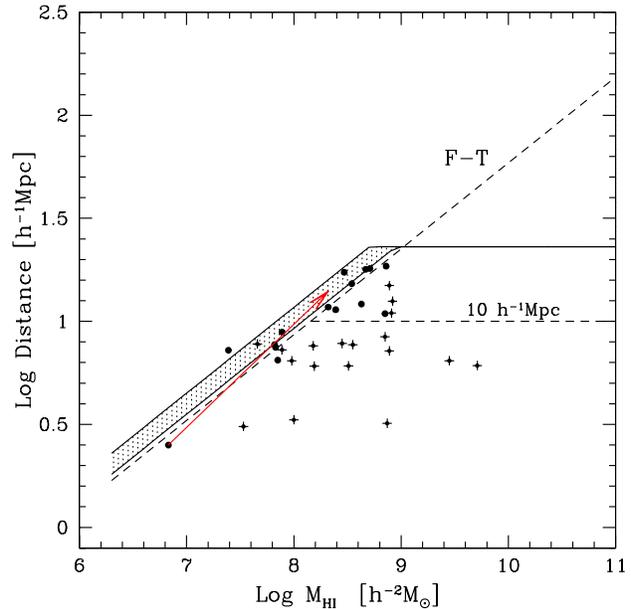,width=8.5cm,angle=0}
  \caption[]{Distance to each of the detected galaxies plotted
as a function of HI-mass, $M_{HI}$. Plus symbols
indicate objects detected by Fisher \& Tully (1981b); solid dots
are for objects that were not. The arrow indicates
the shift implied for the revised distance to
the galaxy UGC~7131 (Makarova \etal 1998).
A diagonal line indicates the sensitivity attained by
the Fisher-Tully Catalog (1981b), as estimated by Briggs (1997).
The cross-hatched band indicates the range of the $4~\sigma$ level, 
depending on the coordinates of the galaxies relative to the
center of the survey strip.}
  \label{d_MHI.fig}
\end{figure}

A convenient plot for comparing relative sensitivities of different
surveys, such as the Fisher-Tully Catalog of Nearby (late-type) Galaxies
(1981b) and the more recent LSB catalogs (Schombert \etal 1997, Sprayberry
\etal 1996) is shown in Fig.~\ref{d_MHI.fig}. Here, the distance to
each galaxy is plotted as a function of its \HI\ mass. Briggs (1997a)
showed that there is a sharp sensitivity boundary to the Fisher-Tully
catalog, indicated by the diagonal dashed line in Fig.~\ref{d_MHI.fig},
and that the newer surveys for LSB galaxies add no substantial number
of objects to the region where Fisher-Tully is sensitive.  The new
objects lie predominantly above the F-T line.  A crucial test 
provided by the new \nan\ survey, is to cover a large area of
sky at a sensitivity matched to the Fisher-Tully sensitivity, to 
determine whether their catalog is indeed complete.  The
result shown in Fig.~\ref{d_MHI.fig} is that the \nan\ survey
finds galaxies both within the F-T zone and above it.  All the
galaxies that we detected within the F-T intended ``zone of 
completeness'' (below their sensitivity line and within 10$h^{-1}$Mpc)
were already included in the F-T Catalog. 
One notable galaxy that was not included in the F-T Catalog, NGC~4203, lies
well within the F-T zone of {\it sensitivity}; it is classified
Hubble type S0 and therefore was not included in Fisher and Tullys'
source list of late-type galaxies. We conclude that the F-T Catalog
is remarkably complete in this region for late-type galaxies, and that 
the biggest incompleteness that may arise when their catalog is
used for measuring the \HI\ content of the nearby universe is that
the F-T Catalog may be lacking the occasional early-type galaxy 
with substantial \HI.

\subsection{The \HI-mass function}
The distribution of \HI-masses is rather homogeneous 
with a mean of log($M_{HI}$) = 8.4 for observed velocities -- and
of log($M_{HI}$) = 8.7 for POTENT corrected HI-masses.
An HI-mass function can be estimated in a straightforward way for
our \nan\ sample. The precision of this computation will be
low for several reasons:  The total number of galaxies is low.
There are no masses below $M_{HI}\approx 3{\times}10^7 M_{\odot}$.
The volume scanned is small, and it cannot be argued that the
sample is drawn from a volume that is respresentative of the
general population, in either \HI\ properties or in the average
number density.  However, the calculation is a useful
illustration of the vulnerability
of these types of calculation to small number statistics and
distance uncertainties.

\begin{figure}
\psfig{figure=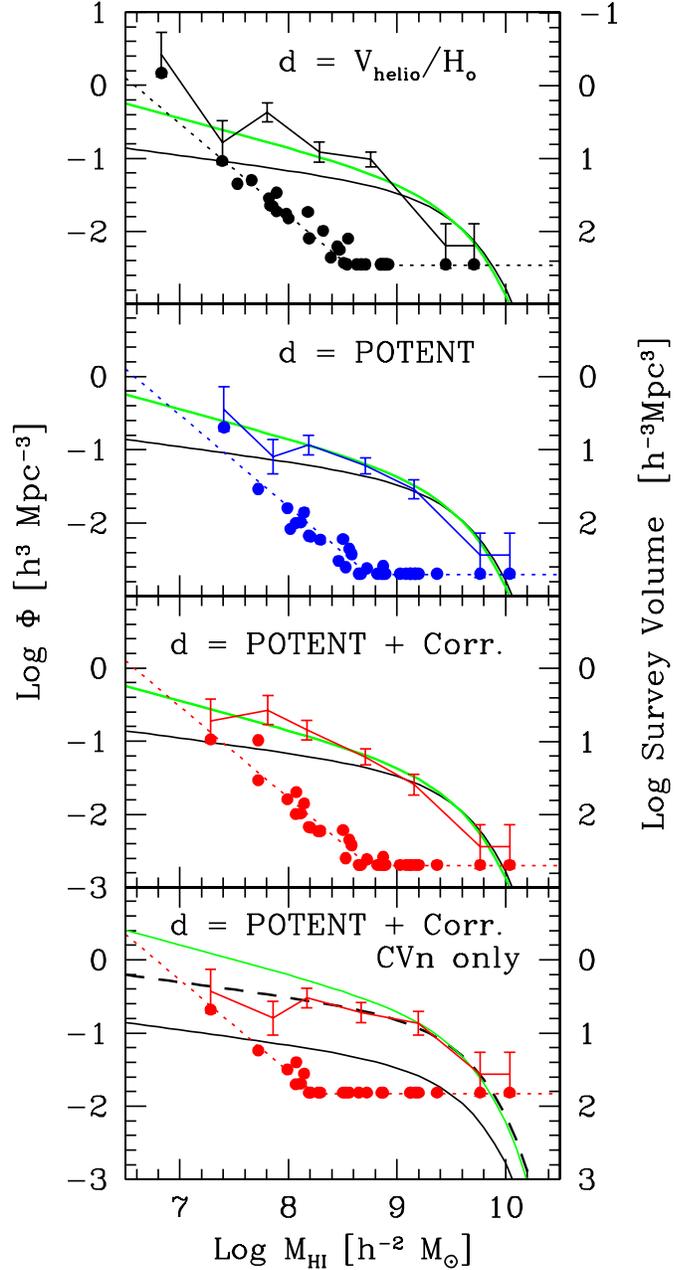,width=8.8cm,angle=0}
  \caption[]{HI-mass function for the Canis Venatici survey
volume, normalized to number of objects per decade of mass.
Error bars represent poisson statistics for the
present sample after binning.  The smooth solid curve is
the analytic form derived by Zwaan \etal (1997) with a slope
of $\alpha =-1.2$, the grey line has a slope of $\alpha=-1.4$
(Banks \etal 1998). The bottom panel
shows the result restricted to the CVn-group regions ($<$1200\kms)
where the dashed curve represents the Zwaan \etal \HI-mass 
function multiplied by a factor of 4.5. The dotted line gives 
an indication of the volume probed as a function of mass 
(see right vertical axis); the points give $1/V_{max}$ 
for each of the galaxies in the sample, taking into account 
the different velocity widths.}
  \label{mfnct_quad.fig}
\end{figure}

We show four derivations of the \HI\ mass function in 
Fig.~\ref{mfnct_quad.fig}.
First, we calculated the number density of galaxies by computing
distances, \HI\ masses, and sensitivity volumes based on 
heliocentric velocities $V_{hel}/H_o$. The mass functions are
binned into half-decade bins, but are scaled to give number of
objects per decade. The value for each decade is computed from 
the sum $\Sigma 1/V_{max}$, where $V_{max}$ is the volume of the
survey in which a galaxy with the properties $M_{HI}$ and $\Delta V$
could have been detected. The values of $1/V_{max}$ are plotted
for every galaxy as dots.  The points representing the number 
density of objects of mass $M_{HI}$ are plotted per bin at
the average $M_{HI}$ for the galaxies included in that bin, so that,
for example, the two highest mass bins, which have only one galaxy each,
are plotted close to each other as upper limits. It is notable
that the galaxy UGC~7131 causes a very steeply rising tail in the
top panel, because
is is treated in this calculation as a very nearby, but low mass
object.  Placed at a greater, more appropriate distance, it becomes
more massive, and it is added to other galaxies of greater velocity
width and higher \HI\ mass in the higher mass bins.

An improved calculation based on the POTENT distances is displayed
in the second panel. In the third panel, the four galaxies
with independent distance measurements have been plotted 
according to their revised distances.
In the 4th panel we have restricted our sample to include only
the overdense foreground region which includes the CVn and Coma
groups, \ie the volume within $V_{hel} < $1200\kms\ and about 
1/2 the RA coverage (about half the solid angle).
Big galaxies can be detected throughout the volume we surveyed, but 
little galaxies can be detected only in the front part of our volume.
The volume normalization factors, which are used to compute the mass
function, are sensitivity limited for the small masses to only the
front part of our survey volume.  For the large masses, the $V_{max}$'s 
include the whole volume, including the volume where the numbers of
galaxies are much less. Hence, when restricting the ``survey volume'' 
we get a fairer comparison of the  number of little galaxies to the 
number of big ones.  

In all four panels the solid line represents the \HI-mass function
with a slope of $\alpha=-1.2$ as derived by Zwaan \etal (1997) from 
the Arecibo blind \HI\ driftscan survey, whereas the grey line represents
an \HI-mass function with a slope of $\alpha=-1.4$ as deduced by
Banks \etal (1998) for a similar but more sensitive survey in the
CenA-group region. 

In the first three panels, the steeper slope
seems to be in closer agreement with the survey results than
the more shallow \HI-mass function with $\alpha=-1.2$.
However, as argued above, the small masses are over-represented in 
comparison to the large masses if we regard the full \nan\ survey
region. This leads to a slope that is too steep for the faint
end. Restricting our volume to the dense foreground region
including ``only'' the CVn and Coma groups, we find that 
the Zwaan \etal\ \HI-mass function with a scaling factor of 4.5
to acount for the local overdensity (dashed line in the bottom panel)
gives an excellent fit to the data.

\section{Conclusion}

The principal conclusion from this survey is that no new HI-rich systems
(LSB galaxies or intergalactic clouds) were discovered.  The previous
deep optical surveys in the CVn group region (BTS), followed by
21cm line observations, have  succeeded in cataloging all the galaxies
that the current \nan\ survey detected. With the follow-up
observations (and their non-confirmations) of all $4~\sigma$ events, the 
survey is complete to \HI\ masses above $8{\times}10^7 M_{\odot}$ 
throughout the CVn Groups. This limit is most strict for the high mass
range. This corroborates the early conclusions made by Fisher and Tully 
in 1981(a) from their blind \HI-survey in the M81-group region with
a limiting sensitivity a factor two higher compared to the here 
performed \HI-survey. 

The logical consequence is that
intergalactic clouds and unseen gas-rich galaxies can form a population
amounting to at most 1/30 the population in optically cataloged large
galaxies. This result is highly consistent with earlier work
(Zwaan \etal 1997, Briggs 1997b, Briggs 1990).

For lower \HI\ masses, the current \nan\ Survey was insufficiently
sensitive for detections throughout the groups.  The depth of the
experiment drops as $d\approx 12(M_{HI}/10^8 \Msun)^{5/12} h^{-1}$kpc,
so that the volume in which small masses ($M_{HI} \la 10^8 M_{\odot}$)
could be detected is small and does not extend to include both CVn Groups.

The \HI-defined sample obtained here, which is complete to 
$8{\times}10^7 M_{\odot}$ throughout the CVn and Coma groups 
is well described by the Zwaan \etal (1997) \HI-mass function  
with a slope of $\alpha= -1.2$ and a scaling factor of 4.5.

\acknowledgements{ 
The Unit\'e Scientifique \nan\
of the Observatoire de Paris is associated as Unit\'e de Service et de
Recherche (USR) No. B704 to the French Centre National de Recherche 
Scientifique (CNRS). The Observatory also gratefully acknowledges the 
financial support of the R\'egion Centre in 
France. While at the Observatoire de Paris (Meudon), the research by 
RCKK was supported with an EC grant.
BB thanks the Swiss National Science Foundation for  financial support.
This research has made use of the NASA/IPAC Extragalactic Database (NED)   
which is operated by the Jet Propulsion Laboratory, California Institute   
of Technology, under contract with the National Aeronautics and Space      
Administration.                                                            
}

\end{document}

%% file: table1.tex
%
%
\begin{table}
\caption{Sky Coverage}
\begin{tabular*}{8cm}{@{\extracolsep\fill}lrrrrrr}
\multicolumn{7}{l}{R.A. : 11$^{h}$ 30$^{m}$--15$^{h}$ 00$^{m}$} \\
  &  &  &  &  &  &  \\
\hline
\vspace{-3 mm} \\
Dec.  &  \multicolumn{6}{c}{No. of cycles} \\    
strip &  &  &  &  &  &  \\
\vspace{-3 mm} \\
\hline
\vspace{-2 mm} \\
35$\dgr$11$'$ & 39 & 38 & 38 & 38 & 33 &  \\   
34$\dgr$49$'$ & 40 & 40 & 40 & 39 & 39 &  \\     
34$\dgr$27$'$ & 40 & 40 & 40 & 37 & 37 &  \\       
34$\dgr$05$'$ & 40 & 40 & 40 & 40 & 40 & 40 \\   
33$\dgr$43$'$ & 40 & 40 & 37 & 37 & 37 &  \\       
33$\dgr$21$'$ & 40 & 40 & 40 & 40 & 38 &  \\        
32$\dgr$59$'$ & 40 & 38 & 38 & 38 & 37 &  \\        
32$\dgr$37$'$ & 40 & 38 & 38 & 38 & 38 &  \\        
32$\dgr$15$'$ & 40 & 40 & 38 & 38 & 38 &  \\         
31$\dgr$53$'$ & 38 & 38 & 38 & 38 &    &  \\           
31$\dgr$31$'$ & 39 & 39 & 38 & 37 & 37 & 36  \\    
31$\dgr$09$'$ & 39 & 39 & 39 & 39 & 39 &  \\        
30$\dgr$47$'$ & 39 & 39 & 39 & 39 & 39 &  \\        
30$\dgr$25$'$ & 39 & 38 & 38 & 38 & 36 &  \\       
30$\dgr$03$'$ & 39 & 37 & 37 & 36 & 35 &  \\     
29$\dgr$41$'$ & 39 & 39 & 37 & 37 & 37 &  \\         
29$\dgr$19$'$ & 39 & 38 & 37 & 35 &    &  \\          
\vspace{-2 mm} \\
\hline
\multicolumn{7}{l}{\footnotesize {\bf Note:} a cycle consists of 20 spectra }\\
\multicolumn{7}{l}{\footnotesize of 16-sec. integration each}\\
\end{tabular*}
\normalsize
\label{skycov}
\end{table}

%% file: table2a.tex
%
%
\begin{table*}
\caption{a -- \HI\ survey detections and possible detections -- 
 reliable detections} 
{\scriptsize
\begin{tabular*}{18cm}{@{\extracolsep\fill}lllrrrrrllllrrlr}
\hline
\vspace{-2 mm} \\
\multicolumn{8}{c}{--- Blind \HI\ line survey data ---} &   \multicolumn{8}{c}{--- Possible optical identifications ---} \\    
No &  R.A. & \hspace{-1 mm}Dec & $V_{\rm HI}$ & $W_{50}$ & $W_{20}$ & $I$ & \MHI\ & Ident. & R.A. & Dec & Classif. &
$D$ & $d$ & $mag$ & $V_{\rm opt}$ \\
 & \multicolumn{2}{c}{(1950.0)} & {\scriptsize km/s} & {\scriptsize km/s} & {\scriptsize km/s} &  
  & \Msun\ &  & \multicolumn{2}{c}{(1950.0)} &  & $'$ & $'$ &  & {\scriptsize km/s} \\ 
\vspace{-3 mm} \\
\hline
\vspace{-2 mm} \\
01 & \hspace{-2 mm}11 36.2 & \hspace{-1 mm}34 05 & 1852 & 180 & 201 &   4.5 & 8.56 &  UGC 6610 & \hspace{-2 mm}11 36 06.0 & \hspace{-1 mm}34 04 58 & Scd:     & 2.1 & 0.4 & 15.0  & 1843 \\  
05 & \hspace{-2 mm}11 41.0 & \hspace{-1 mm}31 53 & 1784 & 143 & 152 &   5.4 & 8.61 &  UGC 6684 & \hspace{-2 mm}11 40 43.9 & \hspace{-1 mm}31 43 59 & Im?      & 0.9 & 0.6 & 15.07 & 1794  \\        
06 & \hspace{-2 mm}11 53.8 & \hspace{-1 mm}31 31 &  668 &  64 &  76 &   2.4 & 7.40 &  BTS 59   & \hspace{-2 mm}11 53 42   & \hspace{-1 mm}31 34 48 & Im/dS0   & 1.2 & 0.6 & 15.5  & \\        
07 & \hspace{-2 mm}11 56.7 & \hspace{-1 mm}30 47 &  764 & 150 & 178 &
8.6 & 8.07 &  NGC 4020 & \hspace{-2 mm}11 56 22.3 & \hspace{-1 mm}30
41 27 & SABd?    & 2.1 & 0.9 & 13.28 &  778  \\        
08 & \hspace{-2 mm}11 59.0 & \hspace{-1 mm}33 43 &  788 &  21 &  63 &   2.4 & 7.54 &  UGC 7007 & \hspace{-2 mm}11 59 00   & \hspace{-1 mm}33 37 10 & Scd/Sm:  & 1.7 & 1.6 & 17    &   \\        
09 & \hspace{-2 mm}12 01.6 & \hspace{-1 mm}32 15 &  743 & 278 & 304 &  24.6 & 8.50 &  NGC 4062 & \hspace{-2 mm}12 01 30.5 & \hspace{-1 mm}32 10 26 & SA(s)c   & 4.1 & 1.7 & 11.9  & 742 \\
10 & \hspace{-2 mm}12 06.9 & \hspace{-1 mm}31 09 &  259 & 106 & 116 &   4.3 & 6.83 &  UGC 7131 & \hspace{-2 mm}12 06 39.3 & \hspace{-1 mm}31 11 04 & Sdm      & 1.5 & 0.4 & 15.10 &   \\         
11 & \hspace{-2 mm}12 06.9 & \hspace{-1 mm}30 03 &  614 &  93 & 109 &  40.0 & 8.55 &  NGC 4136 & \hspace{-2 mm}12 06 45.3 & \hspace{-1 mm}30 12 21 & SAB(r)c  & 4.0 & 3.0 & 11.69 & 445  \\         
   & \hspace{-2 mm}        & \hspace{-1 mm}30 25 &  612 &  81 & 100 &  17.3 &      &           & \hspace{-2 mm}           & \hspace{-1 mm}         &          &     &     &       &   \\  
12 & \hspace{-2 mm}12 10.0 & \hspace{-1 mm}29 19 & 1124 & 164 & 180 &  34.2 & 9.01 &  NGC 4173 & \hspace{-2 mm}12 09 48.4 & \hspace{-1 mm}29 29 18 & Sdm      & 5.0 & 0.7 & 13.59 & 1121  \\        
   & \hspace{-2 mm}        & \hspace{-1 mm}29 41 & 1124 & 164 & 225 &  28.9 &      &           & \hspace{-2 mm}           & \hspace{-1 mm}         &          &     &     &       &   \\ 
13 & \hspace{-2 mm}12 12.4 & \hspace{-1 mm}33 21 & 1093 & 212 & 228 &  27.1 & 8.88 &  NGC 4203 & \hspace{-2 mm}12 12 34   & \hspace{-1 mm}33 28 29 & SAB0     & 3.4 & 3.2 & 11.8  & 1067 \\
14 & \hspace{-2 mm}12 14.4 & \hspace{-1 mm}29 19 & 1210 &  70 &  95 &   3.6 & 8.09 &  UGC 7300 & \hspace{-2 mm}12 14 11.0 & \hspace{-1 mm}29 00 27 & Im       & 1.4 & 1.2 & 14.9  &   \\         
15 & \hspace{-2 mm}12 19.3 & \hspace{-1 mm}35 11 &  732 &  31 &  41 &   2.1 & 7.42 &  UGC 7427 & \hspace{-2 mm}12 19 25.9 & \hspace{-1 mm}35 19 41 & Im       & 1.1 & 0.6 & 16.5  &   \\    
16 & \hspace{-2 mm}12 19.4 & \hspace{-1 mm}32 15 & 1134 &  65 &  72 &   6.3 & 8.28 &  UGC 7428 & \hspace{-2 mm}12 19 31.8 & \hspace{-1 mm}32 22 09 & Im       & 1.3 & 1.2 & 14.1  & 1193 \\  
   & \hspace{-2 mm}        & \hspace{-1 mm}32 27 &\multicolumn{4}{c}{very weak} &  &           & \hspace{-2 mm}           & \hspace{-1 mm}          &          &     &     &       &   \\  
17 & \hspace{-2 mm}12 21.8 & \hspace{-1 mm}31 53 & 1240 & 182 & 203 &  25.7 & 8.97 &  NGC 4359 & \hspace{-2 mm}12 21 41.8 & \hspace{-1 mm}31 47 56 & SB(rs)c? & 3.5 & 0.8 & 13.40 & 1199  \\        
   & \hspace{-2 mm}        & \hspace{-1 mm}31 31 & 1250 & 182 & 201 &   5.4 &      &           & \hspace{-2 mm}           & \hspace{-1 mm}         &          &     &     &       &   \\    
18 & \hspace{-2 mm}12 23.2 & \hspace{-1 mm}33 43 &  310 & 102 & 129 & 389.9 & 8.95 &  NGC 4395 & \hspace{-2 mm}12 23 18   &  \hspace{-1 mm}33 49   & Sd III-IV & 13.2 & 11.0 & 10.64 & 311  \\       
   & \hspace{-2 mm}        & \hspace{-1 mm}34 05 &  326 & 109 & 130 & 175.1 &      &           & \hspace{-2 mm}           &  \hspace{-1 mm}        &          &     &     &       &   \\  
   & \hspace{-2 mm}        & \hspace{-1 mm}33 21 &  303 &  95 & 124 &   9.5 &      &           & \hspace{-2 mm}           &  \hspace{-1 mm}        &          &     &     &       &   \\   
   & \hspace{-2 mm}        & \hspace{-1 mm}34 27 &  304 &  94 & 118 &   4.1 &      &           & \hspace{-2 mm}           & \hspace{-1 mm}         &          &     &     &       &   \\   
19 & \hspace{-2 mm}12 24.0 & \hspace{-1 mm}31 31 &  703 & 331 & 377 &  69.9 & 8.91 &  NGC 4414 & \hspace{-2 mm}12 23 57.9 & \hspace{-1 mm}31 30 00 & SA(rs)c? & 3.6 & 2.0 & 10.96 &  713 \\        
22 & \hspace{-2 mm}12 29.6 & \hspace{-1 mm}30 03 &  654 &  72 &  85 &  11.5 & 8.06 &  UGC 7673 & \hspace{-2 mm}12 29 29.0 & \hspace{-1 mm}29 59 06 & ImIII-IV & 1.4 & 1.3 & 15.28 &   \\  
23 & \hspace{-2 mm}12 30.6 & \hspace{-1 mm}31 53 &  333 &  59 &  74 &  47.3 & 8.09 &  UGC 7698 & \hspace{-2 mm}12 30 26.2 & \hspace{-1 mm}31 49 02 & Im       & 6.5 & 4.5 & 13.0  &   \\           
   & \hspace{-2 mm}        & \hspace{-1 mm}31 31 &  330 &  51 &  73 &  11.5 &      &           & \hspace{-2 mm}           & \hspace{-1 mm}         &          &     &     &       &   \\   
26 & \hspace{-2 mm}12 31.5 & \hspace{-1 mm}30 25 & 1182 & 140 & 148 &   5.2 & 8.23 &  NGC 4525 & \hspace{-2 mm}12 31 23.2 & \hspace{-1 mm}30 33 12 & Scd:     & 2.6 & 1.3 & 12.88 &  1163 \\  
28 & \hspace{-2 mm}12 35.7 & \hspace{-1 mm}33 21 &  331 &  26 &  35 &   2.7 & 7.71 &  UGCA 292 & \hspace{-2 mm}12 36 13   & \hspace{-1 mm}33 02 29 & Im IV-V  & 1.0 & 0.7 & 16.0  &   \\            
   & \hspace{-2 mm}        & \hspace{-1 mm}32 59 &  305 &  29 &  44 &  23.5 &      &           & \hspace{-2 mm}           & \hspace{-1 mm}         &          &     &     &       &   \\     
29 & \hspace{-2 mm}12 37.4 & \hspace{-1 mm}32 59 &  760 &  65 &  72 &   4.4 & 7.78 &  BTS 147  & \hspace{-2 mm}12 37 43.5 & \hspace{-1 mm}32 55 59 & Im       & 1.2 & 0.6 & 15.5  &   \\           
31 & \hspace{-2 mm}12 39.3 & \hspace{-1 mm}32 37 &  604 & 296 & 319 & 604.1 & 9.72 &  NGC 4631 & \hspace{-2 mm}12 39 39.7 & \hspace{-1 mm}32 48 48 & SB(s)d  & 15.5 & 2.7 & 9.75  & 631 \\ 
   & \hspace{-2 mm}        & \hspace{-1 mm}32 59 &  604 & 298 & 317 & 547.3 &      &           & \hspace{-2 mm}           & \hspace{-1 mm}         &          &     &     &       &   \\  
   & \hspace{-2 mm}        & \hspace{-1 mm}33 21 &  581 & 229 & 285 &  35.7 &      &           & \hspace{-2 mm}           & \hspace{-1 mm}         &          &     &     &       &   \\   
32 & \hspace{-2 mm}12 40.9 & \hspace{-1 mm}32 15 &  666 & 101 & 155 & 369.5 & 9.59 &  NGC 4656 & \hspace{-2 mm}12 41 45.4 & \hspace{-1 mm}32 28 43 & SBm      & 15  & 3   & 10.8  & 646  \\           
   & \hspace{-2 mm}        & \hspace{-1 mm}32 27 &  623 & 146 & 200 & 324.5 &      &           & \hspace{-2 mm}           & \hspace{-1 mm}         &          &     &     &       &   \\  
   & \hspace{-2 mm}        & \hspace{-1 mm}32 59 &  637 &  54 & 108 &  37.5 &      &           & \hspace{-2 mm}           & \hspace{-1 mm}         &          &     &     &       &   \\     
33 & \hspace{-2 mm}12 41.7 & \hspace{-1 mm}34 49 &  608 &  58 &  76 &  18.1 & 8.20 &  UGC 7916 & \hspace{-2 mm}12 42 00   & \hspace{-1 mm}34 39 36 & Sm IV    & 3.9 & 2.7 & 14.0  &   \\           
   & \hspace{-2 mm}        & \hspace{-1 mm}34 27 &  605 &  62 &  81 &   9.3 &      &           & \hspace{-2 mm}           & \hspace{-1 mm}         &          &     &     &       &   \\      
34 & \hspace{-2 mm}12 53.7 & \hspace{-1 mm}34 49 &  725 &  54 &  60 &   5.3 & 7.82 &  UGCA 309 & \hspace{-2 mm}12 53 54   & \hspace{-1 mm}34 55 54 & ImIII/N? & 1.6 & 1.4 & 15.0  &   \\    
35 & \hspace{-2 mm}12 56.4 & \hspace{-1 mm}35 11 &  848 &  76 & 107 &  38.6 & 8.82 &  NGC 4861 & \hspace{-2 mm}12 56 38.5 & \hspace{-1 mm}35 06 56 & SB(s)m:  & 6.1 & 2.6 & 12.9  & 810  \\           
   & \hspace{-2 mm}        & \hspace{-1 mm}34 49 &  825 &  54 &  71 &   4.5 &      &           & \hspace{-2 mm}           & \hspace{-1 mm}         &          &     &     &       &   \\   
36 & \hspace{-2 mm}13 02.8 & \hspace{-1 mm}32 59 &  889 &  78 &  96 &   5.4 & 8.00 &  UGC 8181 & \hspace{-2 mm}13 03 02.9 & \hspace{-1 mm}33 10 02 & Sdm      & 1.5 & 0.4 & 16.0  &   \\            
37 & \hspace{-2 mm}13 07.3 & \hspace{-1 mm}34 27 &  811 & 138 & 153 &  19.3 & 8.48 &  UGC 8246 & \hspace{-2 mm}13 07 42   & \hspace{-1 mm}34 27    & SB(s)cd  & 3.5 & 0.6 & 14.6  &   \\           
42 & \hspace{-2 mm}13 37.4 & \hspace{-1 mm}31 31 &  743 &  26 &  51 &   3.1 & 7.60 &  UGC 8647 & \hspace{-2 mm}13 37 30.8 & \hspace{-1 mm}31 32 33 & Im       & 1.1 & 0.3 & 16.5  &   \\          
47 & \hspace{-2 mm}14 37.4 & \hspace{-1 mm}34 27 & 1485 & 185 & 194 &  13.3 & 8.84 &  NGC 5727 & \hspace{-2 mm}14 38 21.3 & \hspace{-1 mm}34 12 07 & SABdm    & 2.2 & 1.2 & 14.2  & 1523  \\        
   & \hspace{-2 mm}        & \hspace{-1 mm}34 05 & 1471 & 163 & 178 &   8.0 &      &           & \hspace{-2 mm}           & \hspace{-1 mm}         &          &     &     &       &   \\  
48 & \hspace{-2 mm}14 43.0 & \hspace{-1 mm}31 25 & 1532 &  98 & 112 &   7.8 & 8.63 &  UGC 9506 & \hspace{-2 mm}14 43 25.4 & \hspace{-1 mm}31 37 33 & Im       & 0.9 & 0.4 & 18    &   \\           
51 & \hspace{-2 mm}14 52.1 & \hspace{-1 mm}31 09 & 1728 & 133 & 157 &   4.9 & 8.54 &  UGC 9597 & \hspace{-2 mm}14 52 53.4 & \hspace{-1 mm}31 01 18 & Sm:      & 1.4 & 1.1 & 17    &   \\         
52 & \hspace{-2 mm}14 53.8 & \hspace{-1 mm}30 25 & 1823 & 106 & 128 &   6.1 & 8.68 &  NGC 5789 & \hspace{-2 mm}14 54 28.8 & \hspace{-1 mm}30 26 08 & Sdm      & 0.9 & 0.8 & 14.17 & 1803  \\ 
\vspace{-2 mm} \\
\hline
\end{tabular*}
}
\label{reldet}
\normalsize
\end{table*}
\normalsize

%% file: table2b.tex
%
%
\addtocounter{table}{-1}
\begin{table*}[t]
\caption{b -- Blind \HI\ survey detections and possible detections -- 
 not confirmed by follow-up pointed observations} 
{\scriptsize
\begin{tabular*}{18cm}{@{\extracolsep\fill}lllrrrrrrllllrrlr}
\hline
\vspace{-2 mm} \\
\multicolumn{9}{c}{--- Blind \HI\ line survey data ---} &   \multicolumn{8}{c}{--- Possible optical identifications ---} \\    
No &  R.A. & Dec & $V_{\rm HI}$ & $W_{50}$ & $W_{20}$ & $I$ & $\sigma$ & \MHI\ & Identif. & R.A. & Dec & Typ &
  $D$ & $d$ & $mag$ & $V_{\rm opt}$ \\
 & \multicolumn{2}{c}{(1950.0)} & {\scriptsize km/s} & {\scriptsize
  km/s} & {\scriptsize km/s} &  &   & \Msun\ &
  & \multicolumn{2}{c}{(1950.0)} &  & $'$ & $'$ &  & {\scriptsize km/s} \\ 
\vspace{-3 mm} \\
\hline
\vspace{-2 mm} \\
02 & \hspace{-2 mm}11 36.2 & 33 21 &  676 &  99 & 106 &  2.6 &  4 & 7.45 &              & \hspace{-2 mm}           &          &         &     &      &      &       \\  
03 & \hspace{-2 mm}11 38.5 & 32 37 & 1835 & 138 & 151 &  5.8 &  8 & 8.66 & Mk 746       & \hspace{-2 mm}11 38 52.5 & 32 37 37 & Im?     & 0.4 & 0.3 & 15.6 &  1758 \\  
   & \hspace{-2 mm}        &       &      &     &     &      &     &      & \hspace{-2 mm}{KUG 1138} & \hspace{-2 mm}11 38 29.5 & 32 42 15 & Im?/Irr & 0.4 & 0.2 & 15.5 &  1735 \\
   & \hspace{-2 mm}        &       &      &     &     &      &     &      & WAS 28       & \hspace{-2 mm}11 38 59.3 & 32 33 30 & \HII\     &     &     & 15.4 &  1768 \\
04 & \hspace{-2 mm}11 40.7 & 29 41 &  380 &  36 &  41 &  2.8 &  7 & 6.98 &              & \hspace{-2 mm}          &          &         &     &     &      &       \\    
20 & \hspace{-2 mm}12 24.2 & 31 53 & 1158 &  21 &  40 &  2.5 & 10 & 7.90 &              & \hspace{-2 mm}           &          &         &     &     &      &       \\  
21 & \hspace{-2 mm}12 28.9 & 31 09 & 1502 &  52 &  59 &  1.3 &  3 & 7.61 &              & \hspace{-2 mm}           &          &         &     &     &      &       \\         
24 & \hspace{-2 mm}12 30.6 & 33 43 &  831 &  24 &  33 &  2.2 &  7 & 7.55 &  MCG 6-29-9  & \hspace{-2 mm}12 30 57   & 33 37 35 & S       & 1.0 & 0.3 & 16   &       \\     
25 & \hspace{-2 mm}12 30.9 & 32 37 & 1868 & 196 & 206 &  2.6 &  3 & 8.33 &              & \hspace{-2 mm}           &          &         &     &     &      &       \\ 
27 & \hspace{-2 mm}12 35.5 & 31 31 &  185 &  64 &  83 &  2.4 &  5 & 6.29 &              & \hspace{-2 mm}           &          &         &     &     &      &       \\        
30 & \hspace{-2 mm}12 38.7 & 32 15 &  489 &  30 &  39 &  2.0 &  6 & 7.05 &              & \hspace{-2 mm}           &          &         &     &     &      &       \\      
38 & \hspace{-2 mm}13 14.5 & 30 03 &  842 &  25 &  35 &  1.8 &  8 & 7.48 &  CG 0999     & \hspace{-2 mm}13 13 55.9 & 30 14 25 & S0$_1$  &     &     & 17.1 &       \\ 
39 & \hspace{-2 mm}13 16.7 & 29 41 & 2217 &  60 &  67 &  4.8 & 10 & 8.74 &              & \hspace{-2 mm}           &          &         &     &     &      &       \\  
40 & \hspace{-2 mm}13 18.5 & 29 19 &  490 &  29 &  61 &  3.8 & 11 & 7.33 &              & \hspace{-2 mm}           &          &         &     &     &      &       \\  
41 & \hspace{-2 mm}13 20.8 & 30 47 & 1679 &  32 &  43 &  2.0 &  6 & 8.12 &              & \hspace{-2 mm}           &          &         &     &     &      &       \\    
43 & \hspace{-2 mm}13 46.0 & 32 59 &  427 &  29 &  39 &  2.9 &  8 & 7.09 &              & \hspace{-2 mm}           &          &         &     &     &      &       \\  
44 & \hspace{-2 mm}13 56.4 & 31 09 &  168 &  26 &  46 &  3.1 &  9 & 6.31 &              & \hspace{-2 mm}           &          &         &     &     &      &       \\  
45 & \hspace{-2 mm}14 14.1 & 31 53 & 1038 &  76 &  92 &  5.7 & 10 & 8.16 &              & \hspace{-2 mm}           &          &         &     &     &      &       \\  
46 & \hspace{-2 mm}14 24.7 & 34 49 & 1157 & 121 & 131 &  7.7 & 11 & 8.38 &              & \hspace{-2 mm}           &          &         &     &     &      &       \\  
   & \hspace{-2 mm}        &   or  & 1201 &  35 &  39 &  3.5 &  9 & 8.07 &              & \hspace{-2 mm}           &          &         &     &     &      &       \\  
49 & \hspace{-2 mm}14 45.7 & 31 31 &  919 &  95 & 118 &  6.1 & 10 & 8.10 &              & \hspace{-2 mm}           &          &         &     &     &      &       \\  
50 & \hspace{-2 mm}14 45.7 & 34 49 &  800 &  37 &  42 &  3.1 &  8 & 7.67 &  UGC 9540?   & \hspace{-2 mm}14 46 48   & 34 55    & triple  & 1.0 & 0.3 & 17   &       \\  
   & \hspace{-2 mm}        & 35 11 &  801 &  30 &  40 &  3.6 & 10 & 7.73 &              & \hspace{-2 mm}           &          &         &     &     &      &       \\  
53 & \hspace{-2 mm}15 05.3 & 33 43 &  171 &  56 &  69 &  2.1 &  4 & 6.16 &              & \hspace{-2 mm}           &          &         &     &     &      &       \\    
\vspace{-2 mm} \\
\hline
\end{tabular*}
\label{uncdet}
}
\normalsize
\end{table*}

%% file: table3.tex
%
%
\begin{table*}
\caption{Pointed observations BTS dwarfs  not detected in the
driftscan survey}
{\scriptsize
\begin{tabular*}{18cm}{@{\extracolsep\fill}lcclllrrllllrcc}
\hline
\vspace{-2 mm} \\
Ident &  R.A. & Dec &  Class & $mag$ & $D$  & Cycl & \multicolumn{2}{c}{rms} 
& $V_{\rm HI}$ & $W_{50}$ & $W_{20}$ &  $I$ &   Ref & Tel \\
      &       &     &        &     &    &      &                Nan & Are \\
 & \multicolumn{2}{c}{(1950.0)} &  &  & $'$ &  & mJy & mJy & {\scriptsize km/s} & {\scriptsize km/s}
  & {\scriptsize km/s} & {\scriptsize Jy km/s} &  &  \\
\vspace{-3 mm} \\
\hline
\vspace{-2 mm} \\     
BTS 060 &  11 54 00 &  30 32 59 & dE,N  & 18.0 & 0.4  & 10 &  3.6 &     \\ 
BTS 113 &  12 15 12 &  33 36 59 & dE/Im & 17.5 & 1.1  & 17 &  3.0 & 1.0 \\
BTS 151 &  12 41 00 &  32 45 00 & dE    & 16.0 & 0.8  & 8  &  4.1 &
& [647] & [45] & [80] & [20.0] & *   & N \\
BTS 156 &  12 46 30 &  32 14 00 & dE    & 17.0 & 0.7  &  8 &  4.6 &     \\ 
BTS 160 &  12 53 00 &  33 15 00 & Im    & 15.5 & 0.7  & 10 &  3.6 &
& 898 & 34 & 58 & 0.9  & H89 & A \\
BTS 161 &  12 53 42 &  34 11 00 & dE/Im & 16.5 & 0.7  & 10 &  3.9 & 1.1 \\
BTS 167 &  14 25 00 &  32 27 00 & Im    & 16.5 & 0.75 & 11 &  3.8 &     \\
\vspace{-2 mm} \\
\hline
\end{tabular*}
}
\label{BTS}
\normalsize
\end{table*}

%% file: table4.tex
%
%
\begin{table*}[ht]
\caption{Comparison of survey data with pointed \HI\ line observations}
\renewcommand{\baselinestretch}{0.8}
{\scriptsize
\begin{tabular*}{18cm}{@{\extracolsep\fill}lllrrrrrrrrrlc}
\hline
\vspace{-1 mm} \\
 &  & \multicolumn{6}{c}{--- \HI\ survey data ---} &   
  \multicolumn{6}{c}{--- pointed \HI\ observations  ---} \\    
No &  Identif. & Dec & $V_{\rm HI}$ & $W_{50}$ & $W_{20}$ & $I$ & \MHI\ & $V_{\rm HI}$ & 
 $W_{50}$ & $W_{20}$ & $I$ & Ref & Tel \\
  &  &  & {\scriptsize km/s} & {\scriptsize km/s} & {\scriptsize km/s} & \hspace{-2 mm}{\scriptsize Jy km/s} &
  \Msun\ & {\scriptsize km/s} &  km/s & {\scriptsize km/s} & \hspace{-2 mm}{\scriptsize Jy km/s} &  &  \\
\vspace{-2 mm} \\
\hline
\vspace{-1 mm} \\
01 & UGC 6610 & 34 05 & 1852 & 180 & 201 &   4.5 &  8.56 & 1851 & 209 & 222 &   8.8 & $\ast$ & N \\ 
   &          &       &      &     &     &       &       & 1851 & 20  & 211 &   9.2 & B92 & N \\
05 & UGC 6684 & 31 53 & 1784 & 143 & 152 &   5.4 &  8.84 & 1788 & 142 & 159 &   5.8 & S90 & A \\   
   &          &       &      &     &     &       &       & 1789 & 108 & 161 &   6.5 & TS79 & G \\       
06 & BTS 59   & 31 31 &  668 &  64 &  76 &   2.4 &  7.40 &  648 &  48 &  65 &   7.1 & H89 & A \\
07 & NGC 4020 & 30 47 &  764 & 150 & 178 &   8.6 &  8.07 &  760 & 162 & 183 &  11.1 & S90 & A \\
   &          &       &      &     &     &       &       &  757 & 175 & 193 &  13.3 & FT81 & E  \\
   &          &       &      &     &     &       &       &  760 &     & 180 &       & G94 & E \\
   &          &       &      &     &     &       &       &  762 & 176 &     &   9.1 & M94 & A \\
08 & UGC 7007 & 33 43 &  788 &  21 &  63 &   2.4 &  7.54 &  771 &  59 &  70 &   2.5 & $\ast$ & N \\ 
   &          &       &      &     &     &       &       &  774 &     &  72 &   3.4 & TC88 & G \\ 
   &          &       &      &     &     &       &       &  786 &     &  67 &   3.8 & FT81 & G \\ 
09 & NGC 4062 & 32 15 &  743 & 278 & 304 &  24.6 &  8.50 &  774 & 303 & 298 &  17.3 & DR78 & G \\ 
   &          &       &      &     &     &       &       &  769 & 312 & 307 &  26.7 & FT81 & G \\ 
   &          &       &      &     &     &       &       &  766 &     & 356 &  24.8 & DS83 & G \\ 
   &          &       &      &     &     &       &       &  765 & 288 & 303 &  19.5 & HS85 & E \\  
   &          &       &      &     &     &       &       &  767 &     & 310 &  23.6 & H83  & A \\ 
   &          &       &      &     &     &       &       &  774 &     &     &       & RD76 & G \\
   &          &       &      &     &     &       &       &  764 & 302 & 303 &  25.1 & BW94 & W \\
   &          &       &      &     &     &       &       &  779 & 325 & 315 &  41.7 & F95 & A \\
10 & UGC 7131 & 31 09 &  259 & 106 & 116 &   4.3 &  6.83 &  253 & 117 & 128 &   4.6 & S90  & A \\ 
   &          &       &      &     &     &       &       &  249 &     &     &       & B85b & A \\
11 & NGC 4136 & 30 03 &  614 &  93 & 109 &  40.0 &  8.85 &  606 &  85 & 104 &  32.1 & $\ast$ & N \\ 
   &          & 30 25 &  612 &  81 & 100 &  17.3 &       &  618 &  94 & 112 &  49.4 & FT81 & G \\ 
   &          &       &      &     &     &       &       &  612 &  89 & 107 &  47.1 & HR86 & E \\  
   &          &       &      &     &     &       &       &  606 &  92 & 109 &  25.0 & L87  & A \\ 
   &          &       &      &     &     &       &       &  596 & 101 & 128 &  35.0 & A79 & Cambr \\
   &          &       &      &     &     &       &       &  607 &     & 105 &       & G94 & E \\
12 & NGC 4173 & 29 19 & 1124 & 164 & 180 &  34.2 &  9.01 & 1122 & 158 & 172 &  33.0 & $\ast$ & N \\  
   &          & 29 41 & 1124 & 164 & 225 &  28.9 &       & 1127 & 170 & 205 &  42.2 & FT81 & G \\
   &          &       &      &     &     &       &       & 1020 &  58 &  98 &  21.6 & WR86 & A \\ 
   &          &       &      &     &     &       &       & 1085 &  83 & 150 &  19.7 & B85  & A \\ 
   &          &       &      &     &     &       &       & 1127 &     & 173 &       & G94  & E \\
13 & NGC 4203 & 33 21 & 1093 & 212 & 228 &  27.1 &  8.88 & 1080 & 230 & 270 &   6.4 & BB77 & A \\ 
   &          &       &      &     &     &       &       & 1083 &     & 264 &  26.7 & BC83 & N \\ 
   &          &       &      &     &     &       &       & 1091 & 240 &     &  27.4 & B87  & A \\ 
   &          &       &      &     &     &       &       & 1093 & 229 & 274 &  20.7 & K77  & B \\
   &          &       &      &     &     &       &       & 1091 &     &     &  27.1 & BK81 & A \\
   &          &       &      &     &     &       &       & 1091 & 240 &     &  27.6 & B87  & A \\
   &          &       &      &     &     &       &       & 1094 &     & 265 &       & G94  & E \\
   &          &       &      &     &     &       &       & 1090 & 243 & 283 &  24   & vD88 & W \\
14 & UGC 7300 & 29 19 & 1210 &  70 &  95 &   3.6 &  8.09 & 1210 &  73 &  91 &  10.4 & S90  & A \\ 
   &          &       &      &     &     &       &       & 1224 &  70 &  96 &       & TW79 & A \\
   &          &       &      &     &     &       &       & 1215 &  78 & 102 &  13.1 & FT81 & G  \\
   &          &       &      &     &     &       &       & 1208 &     &  98 &       & G94 & E  \\
   &          &       &      &     &     &       &       & 1210 &  75 &     &  13.8 & VZ97 & B  \\
15 & UGC 7427 & 35 11 &  732 &  31 &  41 &   2.1 &  7.42 &  719 &  35  & 72 &   0.8 & $\ast$ & N \\    
   &          &       &      &     &     &       &       &  724 &  41 &  61 &   2.8 & S90  & A \\ 
   &          &       &      &     &     &       &       &  725 &  39 &  57 &   2.3 & H89 & A \\
   &          &       &      &     &     &       &       &  729 &     &     &       & B85b & A  \\
16 & UGC 7428 & 32 15 & 1134 &  65 &  72 &   6.3 &  8.28 & 1138 &  67 &  83 &   9.2 & $\ast$ & N \\ 
   &          & 32 37 & \multicolumn{4}{c}{very weak} &  & 1137 &  66 &  85 &   7.3 & S90  & A \\ 
   &          &       &      &     &     &       &       & 1140 &  65 &  83 &   7.6 & L87  & A  \\
17 & NGC 4359 & 31 53 & 1240 & 182 & 203 &  25.7 &  8.97 & 1253 & 204 & 220 &  22.3 & FT81 & E \\ 
   &          & 31 31 & 1250 & 182 & 201 &   5.4 &       &      &     & 216 &       & G94  & E  \\  
18 & NGC 4395 & 33 43 &  310 & 102 & 129 & 389.9 &  8.95 &  320 &     & 140 & 300   & R80  & G  \\ 
   &          & 34 05 &  326 & 109 & 130 & 175.1 &       &  318 & 112 & 132 & 330.2 & HS85 & E \\  
   &          & 33 21 &  303 &  95 & 124 &   9.5 &       &  315 & 101 & 141 & 176.8 & DR78 & G \\  
   &          & 34 27 &  304 &  94 & 118 &   4.1 &       &  318 & 123 & 135 & 334.3 & FT81 & B \\
   &          &       &      &     &     &       &       &  330 & 118 & 135 & 296.7 & W86 & W \\
   &          &       &      &     &     &       &       &  319 &     &     & 405.8 & H98 & B  \\ 
19 & NGC 4414 & 31 31 &  703 & 331 & 377 &  69.9 &  8.91 &  720 & 380 & 418 &  65.1 & FT81 & G \\ 
   &          &       &      &     &     &       &       &  716 & 383 & 422 &  63.1 & RA86 & W  \\
20 & UGC 7673 & 30 03 &  654 &  72 &  85 &  11.5 &  8.06 &  639 &  55 &     &   8.9 & FT75 & G \\ 
   &          &       &      &     &     &       &       &  639 &  58 &  73 &   9.2 & FT81 & G \\ 
   &          &       &      &     &     &       &       &  643 &     &  91 &   9.6 & TC88 & G \\  
   &          &       &      &     &     &       &       &  645 &  70 &     &  14.0 & H96 & A \\
   &          &       &      &     &     &       &       &  649 &  61 &  82 &   6.7 & H89 & A \\
   &          &       &      &     &     &       &       &  644 &     &  86 &       & G94 & E  \\
   &          &       &      &     &     &       &       &  644 &  72 &     &  10.4 & VZ97 & B  \\
   &          &       &      &     &     &       &       &  642 &     &     &       & B85b & A \\
21 & UGC 7698 & 31 53 &  333 &  59 &  74 &  47.3 &  8.09 &  335 &  63 &     &  35.0 & FT75 & G \\ 
   &          & 31 31 &  330 &  51 &  73 &  11.5 &       &  331 &     &  73 &  41.1 & TC88 & G \\  
   &          &       &      &     &     &       &       &  335 &  66 &  93 &  43.1 & FT81 & B \\
   &          &       &      &     &     &       &       &  332 &  55 &  72 &  52.2 & H81  & E \\ 
   &          &       &      &     &     &       &       &  334 &  53 &  72 &  17.9 & H89 & A \\
   &          &       &      &     &     &       &       &  331 &     &     &  41.0 & H98 & B \\
   &          &       &      &     &     &       &       &  334 &  48 &  76 &  29.9 & A79 & Camb \\  
26 & NGC 4525 & 30 25 & 1182 & 140 & 148 &   5.2 &  8.23 & 1165 & 145 & 164 &   6.3 & $\ast$ & N \\
   &          &       &      &     &     &       &       & 1174 &     & 162 &       & G94 & E  \\
   &          &       &      &     &     &       &       & 1177 & 149 & 161 &   6.7 & T98 & N \\ 
   &          &       &      &     &     &       &       & 1172 & 149 &     &   6.5 & K96 & W  \\
28 & UGCA 292 & 32 59 &  305 &  29 &  44 &  23.5 &  7.71 &  309 &  29 &  42 &  13.9 & $\ast$ & N \\   
   &          & 33 21 &  331 &  26 &  35 &   2.7 &       &  312 &     &  51 &  13.6 & FT81 & G \\ 
   &          &       &      &     &     &       &       &  308 &     &  44 &  11.0 & TC88 & G \\ 
   &          &       &      &     &     &       &       &  307 &  25 &     &       & LS79 & E \\ 
   &          &       &      &     &     &       &       &  309 &  30 &     &  20.5 & H96 & A  \\ 
   &          &       &      &     &     &       &       &  308 &  27 &     &  15.7 & VZ97 & B  \\
\vspace{-2 mm} \\
\hline
\end{tabular*}
}
\end{table*}
%
%
\nopagebreak[0]
\addtocounter{table}{-1}
\begin{table*}[t]
\caption{Comparison of survey data with pointed \HI\ line observations 
 -- {\it continued}}
{\scriptsize
\begin{tabular*}{18cm}{@{\extracolsep\fill}lllrrrrrrrrrlc}
\hline
\vspace{-1 mm} \\
 &  & \multicolumn{6}{c}{--- \HI\ survey data ---} &   
  \multicolumn{6}{c}{--- pointed \HI\ observations  ---} \\    
No &  Identif. & Dec & $V_{\rm HI}$ & $W_{50}$ & $W_{20}$ & $I$ & \MHI\ & $V_{\rm HI}$ & 
 $W_{50}$ & $W_{20}$ & $I$ & Ref & Tel \\
  &  &  & {\scriptsize km/s} & {\scriptsize km/s} & {\scriptsize km/s} & \hspace{-2 mm}{\scriptsize Jy km/s} &
  \Msun\ & {\scriptsize km/s} &  km/s & {\scriptsize km/s} & \hspace{-2 mm}{\scriptsize Jy km/s} &  &  \\
\vspace{-2 mm} \\
\hline
\vspace{-1 mm} \\
29 & BTS 147  & 32 59 &  760 &  65 &  72 &   4.4 &  7.78 &  765 &  58 &  90 &   3.8 & H89 & A \\
   &          &       &      &     &     &       &       &  775 &     &     &   5.5 & R94 & W  \\
31 & NGC 4631 & 32 37 &  604 & 296 & 319 & 604.1 &  9.72 &  617 & 286 & 320 & 323.6 & DR78 & G \\  
   &          & 32 59 &  604 & 298 & 317 & 547.3 &       &  613 &     & 320 & 639.3 & R80  & G \\ 
   &          & 33 21 &  582 & 229 & 285 &  35.7 &       &  600 &     &     & 610   & W69  & N \\ 
   &          &       &      &     &     &       &       &  613 & 301 & 325 & 604.3 & FT81 & B \\
   &          &       &      &     &     &       &       &  606 & 261 & 306 & 787.6 & KS77 & A  \\ 
   &          &       &      &     &     &       &       &  606 &     &     &       & KS79 & A  \\ 
   &          &       &      &     &     &       &       &  610 &     & 380 & 506.1 & W78 & W  \\
   &          &       &      &     &     &       &       &  615 &     &     & 428.8 & R94 & W \\
   &          &       &      &     &     &       &       &      &     &     & 766.1 & H75 & E  \\
32 & NGC 4656 & 32 15 &  666 & 101 & 155 & 369.5 &  9.59 &  645 & 126 & 194 & 182.3 & DR78 & G \\  
   &          & 32 27 &  623 & 146 & 200 & 324.5 &       &  644 &     & 174 & 393.0 & R80  & G \\ 
   &          & 32 59 &  637 &  54 & 108 &  37.5 &       &  639 &     & 212 & 305.9 & DS83 & G \\ 
   &          &       &      &     &     &       &       &  646 &     & 184 & 274.6 & TC88 & G \\  
   &          &       &      &     &     &       &       &  649 & 146 & 187 & 327.0 & FT81 & B \\
   &          &       &      &     &     &       &       &  630 &     &     & 258   & W69  & N \\
   &          &       &      &     &     &       &       &  650 &     & 240 & 265.6 & W78 & W  \\ 
   &          &       &      &     &     &       &       &  634 &     &     &       & KS79 & A  \\ 
   &          &       &      &     &     &       &       &  660 &     &     & 302.2 & R94 & W \\
33 & UGC 7916 & 34 49 &  608 &  58 &  76 &  18.1 &  8.20 &  603 &     &  81 &  18.3 & TC88 & G \\  
   &          & 34 27 &  605 &  62 &  81 &   9.3 &       &  612 &  60
   &  87 &  25.4 & FT81 & B  \\   
   &          &       &      &     &     &       &       &  603 &  57 &  77 &  10.5 & H89 & A  \\   
34 & UGCA 309 & 34 49 &  725 &  54 &  60 &   5.3 &  7.82 &  731 &     &  53 &   5.8 & TC88 & G \\  
   &          &       &      &     &     &       &       &  730 &  57 &  67 &   7.0 & FT81 & G \\  
   &          &       &      &     &     &       &       &  717 &  40 &     &   6.4 & LS79 & E \\
   &          &       &      &     &     &       &       &  730 &  42 &     &   5.6 & VZ97 & B  \\  
35 & NGC 4861 & 35 11 &  848 &  76 & 107 &  38.6 &  8.82 &  839 &  89 & 118 &  34.3 & $\ast$ & N \\   
   &          & 34 49 &  825 &  54 &  71 &   4.5 &       &  837 &     & 119 &  39.5 &  C74 & N \\
   &          &       &      &     &     &       &       &  847 &     & 114 &  49.1 & BC81 & N \\
   &          &       &      &     &     &       &       &  847 &  86 & 116 &  40.6 & FT81 & G \\ 
   &          &       &      &     &     &       &       &  843 & 150 & 215 &  56.5 & B82 & N  \\
   &          &       &      &     &     &       &       &  839 &  87 & 110 &  34.1 & BW94 & W  \\
36 & UGC 8181 & 32 59 &  889 &  78 &  96 &   5.4 &  8.00 &  894 & 104 & 146 &   4.4 & $\ast$ & N \\  
   &          &       &      &     &     &       &       &  886 &  97 & 115 &   4.0 & S90  & A \\ 
   &          &       &      &     &     &       &       &  882 &     &     &       & B85b & A \\
37 & UGC 8246 & 34 27 &  811 & 138 & 153 &  19.3 &  8.48 &  813 & 138 & 170 &  19.7 & FT81 & G \\  
   &          &       &      &     &     &       &       &  749 &     &     &       & B85b & A \\
42 & UGC 8647 & 31 31 &  743 &  26 &  51 &   3.1 &  7.60 &  747 &  47 &  77 &   5.1 & S90  & A \\ 
47 & NGC 5727 & 34 27 & 1485 & 185 & 194 &  13.3 &  8.84 & 1489 &     & 217 &  15.1 & H83  & G \\ 
   &          & 34 05 & 1471 & 163 & 178 &   8.0 &       & 1491 & 192 & 196 &  13.6 & FT81 & G \\  
   &          &       &      &     &     &       &       & 1492 &     & 202 &  15.1 & HG84 & A  \\
48 & UGC 9506 & 31 25 & 1532 &  98 & 112 &   7.8 &  8.63 & 1530 &  61 & 119 &   6.9 & $\ast$ & N \\    
   &          &       &      &     &     &       &       & 1519 &  51 & 102 &   4.1 & S90  & A \\
   &          &       &      &     &     &       &       & 1521 &     & 120 &   8.0 & TS79 & G \\ 
51 & UGC 9597 & 31 09 & 1728 & 133 & 157 &   4.9 &  8.80 & 1734 & 135 & 150 &   3.1 & S90  & A \\ 
   &          &       &      &     &     &       &       & 1727 & 128 & 200 &   5.3 & TS79 & G \\
52 & NGC 5789 & 30 25 & 1823 & 106 & 128 &   6.1 &  8.68 & 1800 & 118 & 190 &   8.0 & P79 & G \\ 
   &          &       &      &     &     &       &       & 1801 &     & 165 &   6.9 & TC88 & G \\  
   &          &       &      &     &     &       &       & 1806 &  96 & 133 &   5.8& OS93 & W \\
   &          &       &      &     &     &       &       & 1805 & 105 &     &   5.9 & C93 & A \\
\vspace{-2 mm} \\
\hline
\end{tabular*}
}
\nopagebreak[0]
\vspace*{-3mm}
\normalsize
%
\addtocounter{table}{-1}
\caption{References and Telescope Codes}
{\scriptsize
\begin{tabular*}{18cm}{@{\extracolsep\fill}llllll}
\vspace{-1 mm} \\
A79   & Allsopp (1979)                 & 
BC81  & Balkowski \& Chamaraux (1981)  & 
BC83  & Balkowski \& Chamaraux (1983)  \\
BB77  & Bieging \& Biermann (1977)     & 
B85   & Bothun et al. (1985)           & 
B85b  & Bothun et al. (1985b)          \\ 
B82   & Bottinelli et al. (1982)       &
B92   & Bottinelli et al. (1992)       &
BW94  & Broeils \& van Woerden (1994)  \\
BK81  & Burstein \& Krumm (1981)       &
B87   & Burstein et al. (1987)         & 
C74   & Carozzi et al. (1974)          \\ 
C93   & Chengalur et al. (1993)        &
DS83  & Davis \& Seaquist (1983)       &
DR78  & Dickel \& Rood (1978)          \\ 
FT75  & Fisher \& Tully (1975)         & 
FT81  & Fisher \& Tully (1981)         & 
F95   & Freudling (1995)               \\
G94   & Garcia-Baretto et al. (1994)   &
HG84  & Haynes \& Giovanelli (1984)    &
H98   & Haynes et al. (1998)           \\  
H83   & Hewitt et al. (1983)           & 
H89   & Hoffman et al. (1989)          &            
H96   & Hoffman et al. (1996)          \\
H75   & Huchtmeier (1975)              &
H81   & Huchtmeier et al. (1981)       &    
HS85  & Huchtmeier \& Seiradakis (1985)\\  
HR86  & Huchtmeier \& Richter (1986)   &
K96   & Kamphuis et al. (1996)          & 
K77   & Knapp et al. (1977)            \\
KS77  & Krumm \& Salpeter (1977)       &
KS79  & Krumm \& Salpeter (1979)       &   
L87   & Lewis (1987)                   \\
LS79  & Lo \& Sargent (1979)           &
M94   & Magri (1994)                   &  
OS93  & Oosterloo \& Shostak (1993)    \\
P79   & Peterson (1979)                &
R94   & Rand (1994)                    &  
RA86  & Rhee \& van Albada (1996)      \\
RD76  & Rood \& Dickel (1976)          &
R80   & Rots (1980)                    &  
S90   & Schneider et al. (1990)        \\
TW79  & Tarter \& Wright (1979)        &
T98   & Theureau et al. (1998)         &  
TS79  & Thuan \& Seitzer (1979)        \\
TM81  & Thuan \& Martin (1981)         &
TC88  & Tifft \& Cocke (1988)          &  
vD88  & van Driel et al. (1988)        \\
VZ97  & Van Zee et al. (1997)          &
W69   & Weliachew (1969)               &  
W78   & Weliachew et al. (1978)        \\
W86   & Wevers et al. (1986)           &
WR86  & Williams \& Rood (1986)        &  
$\ast$  & this paper                   \\ 
\vspace{-3 mm} \\
\hline 
\vspace{-2 mm} \\
A  & Arecibo  & 
B  & Green Bank 43-m.  & 
E  & Effelsberg \\ 
G  & Green Bank 90-m.  & 
N  & Nancay \\ 
\vspace{-3 mm} \\
\hline
\end{tabular*}
}
\end{table*} 
\renewcommand{\baselinestretch}{1.0}
\normalsize

%% file: table5.tex
%
%
\begin{table*}
\caption{\HI\ properties of galaxies
detected in the drift-scan CVn galaxy search}
{\scriptsize
\begin{tabular*}{18cm}{@{\extracolsep\fill}llllrrrrrrrrrrc}
\hline
\vspace{-2 mm} \\
No &  Ident. & Class. & $i$ & $W_{50}$ & $W_{20}$ & $W_{50}^{\rm cor}$ &
  $W_{20}^{\rm cor}$ & $V_{\rm HI}$ & \MHI\ & $M_{\rm B}$ & 
$V_{\rm HI}^{\rm POT} $ & \MHI \hspace{-3mm}$^{\rm POT}$ & $M_{\rm B}^{\rm POT}$ & {\scriptsize \MHILB} \\
 &  &  &  \dgr\ & {\scriptsize km/s} & {\scriptsize km/s} & {\scriptsize km/s} & {\scriptsize km/s} &   
  {\scriptsize km/s} & \Msun\ &  mag & {\scriptsize km/s} & \Msun\ &  mag & {\scriptsize \MsunLBsun} \\
\vspace{-3 mm} \\
\hline
\vspace{-2 mm} \\
 01 & UGC 6610 & Scd:      & 79 & 207 & 216 & 210 & 220 & 1851 &  8.86 & -16.3 & 2396 &  9.08 & -16.5 &  1.41 \\
 05 & UGC 6684 & Im?       & 48 & 125 & 160 & 168 & 215 & 1789 &  8.67 & -16.1 & 2316 &  8.89 & -16.3 &  1.09 \\
 06 & BTS 59   & Im/dS0    & 60 &  48 &  65 &  55 &  75 &  648 &  7.85 & -13.6 &  976 &  8.21 & -14.0 &  1.65 \\
 07 & NGC 4020 & SABd?     & 65 & 171 & 185 & 188 & 204 &  760 &  8.18 & -16.1 & 1103 &  8.50 & -16.4 &  0.35 \\
 08 & UGC 7007 & Scd/Sm:   & 20 &  59 &  70 & 172 & 205 &  777 &  7.66 & -12.4 & 1137 &  7.99 & -12.7 &  3.22 \\
 09 & NGC 4062 & SA(s)c    & 65 & 306 & 313 & 337 & 345 &  770 &  8.55 & -17.4 & 1118 &  8.87 & -17.7 &  0.25 \\
 10 & UGC 7131 & Sdm       & 74 & 117 & 128 & 121 & 133 &  251 &  6.83 & -11.9 &  487 &  7.41 & -12.5 &  0.76 \\
 11 & NGC 4136 & SAB(r)c   & 41 &  92 & 111 & 140 & 169 &  608 &  8.51 & -17.2 &  906 &  8.86 & -17.5 &  0.27 \\
 12 & NGC 4173 & Sdm       & 82 & 117 & 160 & 118 & 162 & 1096 &  8.91 & -16.6 & 1479 &  9.17 & -16.9 &  1.20 \\
 13 & NGC 4203 & SAB0      & 20 & 236 & 271 & 690 & 792 & 1089 &  8.85 & -18.3 & 1489 &  9.12 & -18.6 &  0.22 \\
 14 & UGC 7300 & Im        & 31 &  74 &  97 & 143 & 188 & 1213 &  8.63 & -15.5 & 1607 &  8.87 & -15.7 &  1.73 \\
 15 & UGC 7427 & Im        & 57 &  38 &  63 &  45 &  75 &  724 &  7.39 & -12.8 & 1061 &  7.72 & -13.1 &  1.20 \\
 16 & UGC 7428 & Im        & 23 &  66 &  84 & 168 & 215 & 1138 &  8.39 & -16.1 & 1531 &  8.65 & -16.4 &  0.57 \\
 17 & NGC 4359 & SB(rs)c?  & 23 & 204 & 218 & 522 & 558 & 1253 &  8.92 & -17.0 & 1658 &  9.16 & -17.2 &  0.85 \\
 18 & NGC 4395 & Sd III-IV & 34 & 113 & 137 & 202 & 245 &  320 &  8.87 & -16.8 &  569 &  9.37 & -17.3 &  0.91 \\
 19 & NGC 4414 & SA(rs)c?  & 56 & 382 & 420 & 460 & 507 &  718 &  8.89 & -18.2 & 1027 &  9.20 & -18.5 &  0.26 \\
 22 & UGC 7673 & ImIII-IV  & 22 &  63 &  83 & 168 & 222 &  643 &  7.98 & -13.7 &  923 &  8.29 & -14.0 &  2.03 \\
 23 & UGC 7698 & Im        & 46 &  57 &  77 &  79 & 107 &  333 &  8.00 & -14.6 &  567 &  8.46 & -15.1 &  0.93 \\
 26 & NGC 4525 & Scd:      & 60 & 148 & 162 & 170 & 187 & 1172 &  8.32 & -17.4 & 1545 &  8.56 & -17.6 &  0.15 \\
 28 & UGCA 292 & Im IV-V   & 46 &  28 &  46 &  38 &  64 &  309 &  7.53 & -11.4 &  542 &  8.02 & -11.9 &  6.00 \\
 29 & BTS 147  & Im        & 60 &  58 &  90 &  66 & 104 &  770 &  7.82 & -13.9 & 1082 &  8.12 & -14.2 &  1.17 \\
 31 & NGC 4631 & SB(s)d    & 80 & 283 & 330 & 287 & 335 &  610 &  9.71 & -19.1 &  891 & 10.04 & -19.4 &  0.76 \\
 32 & NGC 4656 & SBm       & 78 & 136 & 199 & 139 & 203 &  644 &  9.45 & -18.2 &  927 &  9.77 & -18.5 &  0.95 \\
 33 & UGC 7916 & Sm IV     & 46 &  59 &  82 &  82 & 114 &  606 &  8.19 & -14.9 &  894 &  8.53 & -15.2 &  1.09 \\
 34 & UGCA 309 & ImIII/N?  & 29 &  46 &  60 &  94 & 124 &  727 &  7.89 & -14.3 & 1025 &  8.19 & -14.6 &  0.95 \\
 35 & NGC 4861 & SB(s)m:   & 65 & 103 & 132 & 113 & 146 &  842 &  8.85 & -16.7 & 1157 &  9.13 & -17.0 &  0.95 \\
 36 & UGC 8181 & Sdm       & 74 & 101 & 131 & 105 & 136 &  887 &  7.89 & -13.7 & 1190 &  8.15 & -14.0 &  1.65 \\
 37 & UGC 8246 & SB(s)cd   & 80 & 138 & 170 & 140 & 173 &  781 &  8.45 & -14.9 & 1070 &  8.72 & -15.2 &  1.99 \\
 42 & UGC 8647 & Im        & 74 &  45 &  77 &  46 &  80 &  747 &  7.83 & -12.8 &  982 &  8.07 & -13.0 &  3.30 \\
 47 & NGC 5727 & SABdm     & 57 & 192 & 208 & 228 & 248 & 1491 &  8.89 & -16.6 & 1748 &  9.03 & -16.7 &  1.14 \\
 48 & UGC 9506 & Im        & 64 &  56 & 114 &  62 & 127 & 1523 &  8.54 & -12.9 & 1767 &  8.67 & -13.0 & 15.42 \\
 51 & UGC 9597 & Sm:       & 38 & 132 & 175 & 214 & 284 & 1730 &  8.47 & -14.1 & 1973 &  8.58 & -14.2 &  4.35 \\
 52 & NGC 5789 & Sdm       & 27 & 106 & 163 & 233 & 359 & 1803 &  8.71 & -17.1 & 2047 &  8.82 & -17.2 &  0.48 \\
\vspace{-2 mm} \\
\hline
\multicolumn{12}{l}{\footnotesize {\bf Note:} For comments on the inclination of the \HI\ 
 gas in NGC 4203 (no. 15), see Sect 4.3} \\
\end{tabular*}
}
\label{globprop}
\normalsize
\end{table*}

%% file: cvn.bbl
\begin{thebibliography}{}
\bibitem[]{}
Allsopp N.J. 1979, MNRAS 188, 371
\bibitem[]{}
Balkowski C., Chamaraux P. 1981, A\&A 97, 223 
\bibitem[]{}
Balkowski C., Chamaraux P. 1983, A\&AS 51, 331 
\bibitem[]{}
Banks G.D., Disney M.J., Knezek P., \etal 1998, in prep
\bibitem[]{}
Bertschinger E., Dekel A., Faber S.M., Dressler A., Burstein D.
1990, ApJ, 364, 370
\bibitem[]{}
Bieging J.H., Biermann P. 1977, A\&A 26, 361
\bibitem[]{}
Binggeli B., Tarenghi M., Sandage A. 1990, A\&A 228, 42 (BTS)
\bibitem[]{}
Bothun G.D., Aaronson M., Schommer R., \etal 1985, ApJS 57, 423
\bibitem[]{}
Bothun G.D., Beers T.C., Mould J.R. 1985b, AJ 90, 2487
\bibitem[]{}
Bottinelli L., Gouguenheim L., Paturel G. 1982, A\&A 113, 61
\bibitem[]{}
Bottinelli L., Durand N., Fouque P., \etal 1992, A\&AS 93,173 
\bibitem[]{}
Briggs F.H. 1990, AJ, 100, 999
\bibitem[]{}
Briggs F.H. 1997a, ApJ, 484, L29 
\bibitem[]{}
Briggs F.H. 1997b, ApJ, 484, 618 
\bibitem[]{}
Briggs F.H., Rao S. 1993, ApJ, 417, 494
\bibitem[]{}
Briggs F.H., Sorar E., Kraan-Korteweg R.C., van Driel, W. 1997, PASA 14, 37
\bibitem[]{}
Broeils A.H., van Woerden H. 1994, A\&AS 107, 129   
\bibitem[]{}
Burstein D., Krumm N. 1981, ApJ 250, 517  
\bibitem[]{}
Burstein D., Krumm N., Salpeter E.E. 1987, AJ 94, 883
\bibitem[]{}
Carozzi N., Chamaraux P., Duflot-Augarde R. 1974, A\&A 30, 21
\bibitem[]{}
Chengalur J.N., Salpeter E.E., Terzian Y. 1993, ApJ 419, 30 
\bibitem[]{}
Dalcanton J.J., Spergel D.N., Gunn J.E., Schmidt M., Schneider
D.P. 1997, AJ 114, 635
\bibitem[]{}
Davis L.E., Seaquist E.R. 1983, ApJS 53, 269
\bibitem[]{}
De Vaucouleurs G. 1975, in: Stars and Stellar Systems. IX. Galaxies and the
     Universe, eds. A.Sandage, M.Sandage, and J.Kristian (Chicago University
     Press, Chicago), p.557
\bibitem[]{}
Dickel J.R. Rood H.J. 1978, ApJ 223, 391
\bibitem[]{}
Ferguson H.C., Binggeli B. 1994, A\&A Rev 6, 67
\bibitem[]{}
Fisher J.R., Tully R.B. 1975, A\&A 44, 151
\bibitem[]{}
Fisher J.R., Tully R.B. 1981a, ApJ 243, L23
\bibitem[]{}
Fisher J.R., Tully R.B. 1981b, ApJS 47, 139
\bibitem[]{}
Freudling W. 1995, A\&AS 112, 429 
\bibitem[]{}
Garcia-Baretto J.A., Downes D., Huchtmeier W.K. 1994, A\&A 288, 705
\bibitem[]{}
Haynes M.P., Giovanelli R. 1984, AJ 89, 758
\bibitem[]{}
Haynes M.P., Hogg D.E., Maddalena R.J., Roberts M.S., Van Zee L. 1998, AJ 115, 62 
\bibitem[]{}
Hewitt J.N., Haynes M.P., Giovanelli R. 1983, AJ 88, 272
\bibitem[]{}
Hoffman G.L., Helou G., Salpeter E.E. 1988, ApJ, 324, 75
\bibitem[]{}
Hoffman G.L., Salpeter E.E., Farhat B., \etal 1996, ApJS 105, 269 
\bibitem[]{}
Hoffman G.L., Williams H.L., Salpeter E.E., Sandage A., Binggeli B. 1989, ApJS 71, 701
\bibitem[]{}
Huchtmeier W.K. 1975, A\&A 45, 259        
\bibitem[]{}
Huchtmeier W.K., Seiradakis J.H., Materne J. 1981, A\&A 102, 134
\bibitem[]{}
Huchtmeier W.K., Seiradakis J.H. 1985, A\&A 143, 216
\bibitem[]{}
Huchtmeier W.K., Richter O.-G. 1986, A\&AS 63, 323
\bibitem[]{}
Impey C.D., Sprayberry D., Irwin M.J., Bothun, G.D. 1996, ApJS 105, 2091
\bibitem[]{}
Kamphuis J.J., Sijbring D., van Albada T.S. 1996, A\&AS 116, 15
\bibitem[]{}
Karachentsev I.D., Drozdovsky, I.O. 1998,  subm to A\&AS
\bibitem[]{}
Knapp G.R., Gallagher J.S., Faber S.M., Balick B. 1977, AJ 82, 106 
\bibitem[]{}
Kraan-Korteweg R.C. 1986, A\&AS 66, 255
\bibitem[]{}
Krumm N., Salpeter E.E. 1977, A\&A 56, 465
\bibitem[]{}
Krumm N., Salpeter E.E. 1979, AJ 84, 1138 
\bibitem[]{}
Lewis B.M. 1987, ApJS 63, 515
\bibitem[]{}
Lo K.Y., Sargent W.L.W. 1979, ApJ 227, 756
\bibitem[]{}
Magri C. 1994, AJ 108, 896 
\bibitem[]{}
 Makarova L., Karachentsev I., Takalo L., Heinamaki P.,
 Valtonen M. 1998, subm to A\&AS  
\bibitem[]{}
Matthews L.D., van Driel W., Gallagher J.S. III. 1998, A\&A, in press
\bibitem[]{}
Oosterloo T., Shostak G.S. 1993, A\&AS 99, 379
\bibitem[]{}
Peterson S.D. 1979 ApJ 232, 20 or ApJS 40, 527
\bibitem[]{}
Rand R.J. 1994, A\&A 285, 833 
\bibitem[]{}
Rhee G., van Albada T.S. 1996, A\&AS 115, 407
\bibitem[]{}
Rood H.J., Dickel J.R. 1976, ApJ 205, 346 
\bibitem[]{}
Rots A.H. 1980, A\&AS 41, 189
\bibitem[]{}
Sandage A., Binggeli B. 1984, AJ, 89, 919
\bibitem[]{}
Schneider S.E., Thuan T.X., Magri C., Wadjak J.E. 1990, ApJS 72, 245
\bibitem[]{}
Schombert J.M., Pildis R.A., Eder J.A. 1997, ApJS 111, 233
\bibitem[]{}
Sorar E. 1994, Ph.D. Thesis, University of Pittsburgh
\bibitem[]{}
Sprayberry D., Impey C.D., Irwin M.J. 1996, ApJ 463, 535
\bibitem[]{}
Staveley-Smith L. 1985, Ph.D. Thesis, Univ. of Manchester 
\bibitem[]{}
Tarter J.C., Wright M.C.H. 1979, A\&A 76, 127
\bibitem[]{}
Theureau G., Bottinelli L., Coudreau-Durand N., \etal 1998, A\&AS, in
press
\bibitem[]{}
Theureau G., Bottinelli L., Gouguenheim L. 1997, A\&AS in press
\bibitem[]{}
Thuan T.X., Martin G.E. 1981, ApJ 247, 823
\bibitem[]{}
Thuan T.X., Seitzer P.O. 1979, ApJ 231, 327  
\bibitem[]{}
Tifft W.G., Cocke W.J. 1988, ApJS 67,1
\bibitem[]{}
Tully R.B.,  Fisher J.R. 1987, Nearby Galaxies Atlas (Cambridge University
     Press, Cambridge)
\bibitem[]{}
van Driel W., van Woerden H., Gallagher J.S. 1988 A\&A 191, 201
\bibitem[]{}
Van Zee L., Maddalena R.J., Haynes M.P., Hogg D.E., Roberts M.S. 1997, AJ 113 1638
\bibitem[]{}
Weliachew L. 1969, A\&A 24, 59    
\bibitem[]{}
Weliachew L., Sancisi R., Guelin M. 1978, A\&A 65, 37
\bibitem[]{}
Wevers B.M.H.R., van der Kruit P.C., Allen R.J. 1986, A\&AS 66, 505
\bibitem[]{}
Williams B.A., Rood H.J. 1986, ApJS 63, 265
\bibitem[]{}
Zwaan M.A., Briggs F.H., Sprayberry D., Sorar E. 1997, ApJ 490, 173
\bibitem[]{}
Zwaan M.A., Dalcanton J., Verheijen M., Briggs F. 1998, in prep.
\end{thebibliography}
